\documentclass[12pt]{article}
\usepackage{amsmath,amsthm,amssymb,euscript,epsf,epsfig,color,graphicx}
\usepackage{array}
\usepackage{fancybox}

\usepackage{comment} 
\usepackage[utf8]{inputenc}
\usepackage[english]{babel}
\usepackage{hyperref}

\newcommand{\be}{\begin{equation}}
\newcommand{\ee}{\end{equation}}
\newcommand{\bea}{\begin{eqnarray}\displaystyle}
\newcommand{\eea}{\end{eqnarray}}
\newcommand{\nnm}{\nonumber}

\makeatletter
\@addtoreset{equation}{section}
\makeatother
\renewcommand{\theequation}{\thesection.\arabic{equation}}
\def\one{{\hbox{ 1\kern-.8mm l}}}
\def\zero{{\hbox{ 0\kern-1.5mm 0}}}

\def\mR{ \mathbb{R}} 
\def\mC{ \mathbb{C}} 
\def\tr{ \rm{tr}}
 
\def\Dim{ {\rm Dim} }

\def\con{ \rm{conn} }

\def\cM{{\cal M}} \def\cN{{\cal N}} 
  
\def\cS{{\cal S}}  
  
 \def\cZ{{\cal Z}}

\def\nat{ \rm{nat} }

\def\Sym{ {\rm{Sym} }}

\begin{document}

\makeatletter
\@addtoreset{equation}{section}
\makeatother
\renewcommand{\theequation}{\thesection.\arabic{equation}}


\makeatletter
\@addtoreset{equation}{section}
\makeatother
\renewcommand{\theequation}{\thesection.\arabic{equation}}
\def\ls[#1]{ {}_{#1}}

\def\wmu{ \widetilde{\mu} } 
\def\wmuo{ \widetilde{\mu}_{ \ls[1] } }
\def\wmut{  \widetilde{\mu}_{ \ls[2] } }

\rightline{QMUL-PH-18-17}
\vspace{1.8truecm}

\vspace{10pt}

{\LARGE{ 
\centerline{\bf  Permutation Invariant Gaussian  Matrix Models } 
}}  

\vskip.5cm 

\thispagestyle{empty} \centerline{
   {\large \bf  Sanjaye Ramgoolam${}^{a,b,}$\footnote{ {\tt s.ramgoolam@qmul.ac.uk}},    }}

\vspace{.2cm}
\centerline{{\it ${}^a$ Centre for Research in String Theory, School of Physics and Astronomy},}
\centerline{{ \it Queen Mary University of London},} \centerline{{\it
    Mile End Road, London E1 4NS, UK}}
    
    \vspace{.2cm}
\centerline{{\it ${}^b$ National Institute for Theoretical Physics,}}
\centerline{{\it School of Physics and Centre for Theoretical Physics, }}
\centerline{{\it University of the Witwatersrand, Wits, 2050, South Africa } }

\vspace{.5truecm}

\thispagestyle{empty}

\centerline{\bf ABSTRACT}

\vskip.2cm 

Permutation invariant Gaussian matrix models were recently developed for applications in computational linguistics. A 5-parameter family of models was solved. In this paper, we use a representation theoretic approach to solve the general 13-parameter Gaussian model, which can be viewed as a zero-dimensional quantum field theory.  We express  the  two linear and eleven  quadratic terms in the action in terms of representation theoretic parameters. These parameters  are coefficients of simple  quadratic expressions in terms of appropriate linear combinations of the matrix variables transforming in specific irreducible representations of the symmetric group $S_D$ where $D$ is the size of the matrices. They allow  the identification of constraints which ensure a convergent Gaussian measure and well-defined expectation values for  polynomial functions of the random matrix at  all orders. A graph-theoretic interpretation is known to allow the  enumeration of permutation invariants of matrices at linear, quadratic and higher orders. We express the expectation values of all the quadratic graph-basis invariants and a selection of cubic and quartic invariants in terms of the representation theoretic parameters of the model.

\setcounter{page}{0}
\setcounter{tocdepth}{2}

\newpage

\tableofcontents

\section{ Introduction}

In the context of distributional semantics \cite{Harris,Firth}, the meaning of words is represented 
by vectors which are constructed from the co-occurrences of a word of interest with a set of context words.  In tensorial compositional distributional semantics \cite{CSC2010,GS2015,MCG2014,BBZ2014,KSP2012}, different types of words, depending on  their grammatical role, are associated with vectors, matrices or higher rank tensors.  In \cite{LMT,LMTqpl} we initiated a study of the statistics of these tensors in the framework of  matrix/tensor models. We focused on matrices associated with adjectives or verbs, constructed  by a linear regression method, from the vectors for nouns and for adjective-noun composites or verb-noun composites.  
  
We developed a 5-parameter Gaussian model, 
  \bea\label{themodel}  
  \cZ ( \Lambda ,  a, b , J^0 , J^S  )&&  = \int dM e^{ - { \Lambda \over 2 } \sum_{ i = 1 }^D   M_{ii}^2  - { 1 \over 4 } ( a + b )  \sum_{ i < j } 
  ( M_{ij}^2  + M_{ji}^2 ) } \cr 
  && \hskip1.5cm e^{ -  { 1 \over 2 } ( a - b ) \sum_{ i < j } M_{ij} M_{ji}  +  J^0  \sum_{ i} M_{ii} + J^S  \sum_{ i <  j }  ( M_{ij} + M_{ ji } )  }  \, . 
\eea
The parameters $ J^S , J^0 , a , b , \Lambda $ are coefficients of five linearly independent 
linear and quadratic functions of the $D^2$ random matrix variables $M_{ i , j } $ which are permutation invariant, i.e. obey the equation 
\bea\label{perminv} 
f ( M_{ i , j }  ) = f ( M_{ \sigma (i) , \sigma (j) } ) 
\eea
for $ \sigma \in S_D $, the symmetric group of all permutations of $D$ distinct objects. This $S_D$ invariance implements the notion that the meaning represented by the word-matrices 
is independent of  the ordering of the $D$ context words.   General observables of the model are polynomials $f(M)$ obeying the condition (\ref{perminv}). 
At quadratic order there are $11$  linearly independent polynomials, which are listed in Appendix B
of \cite{LMT}. 
A three dimensional subspace of  quadratic invariants was used in the model above. 
 The most general Gaussian  matrix model compatible with  $ S_D$ symmetry 
considers all the eleven parameters and allows coefficients for each of them. 
What makes the 5-parameter model relatively easy  to handle is that the diagonal variables $M_{ii}$ are each decoupled from each other and from the off-diagonal elements, and there are $ D ( D-1)/2$ pairs of 
off-diagonal elements. For each $i < j$, $M_{ij} $ and $M_{ ji}$ mix with each other so the solution of the model requires an inversion of a $ 2 \times 2 $ matrix.

Expectation values of $f ( M ) $ are computed as 
\bea 
\langle f ( M ) \rangle \equiv { 1 \over \cZ } \int dM f ( M ) \text{EXP}
\eea
where EXP is the product of  exponentials  in (\ref{themodel}).  

Representation theory of $S_D$ offers the techniques to solve
the general permutation invariant   
Gaussian model.  The $D^2$ matrix elements $M_{ ij}$ transform as the tensor product  
$V_D \otimes V_D$ of two copies of the natural representation $V_D$. 
We first decompose $V_D \otimes V_D$ into irreducible representations of 
the diagonal $S_D$. 
\bea 
V_D \otimes V_D =  2 V_0 \oplus  3 V_H \oplus V_2 \oplus V_3
\eea
The trivial (one-dimensional) representation $ V_0$ occurs with multiplicity $2$. 
The $(D-1)$-dimensional irreducible representation (irrep)   $V_H$ occurs with multiplicity $3$. 
$V_2$ is  an irrep of dimension $ { ( D-1)(D-3) \over 2 } $ which occurs with multiplicity $1$. 
Likewise, $V_3$ of dimension $ { ( D-1)( D-2) \over 2 } $ occurs with multiplicity $1$. 
As a result of these multiplicities, the 11 parameters can be decomposed as 
\bea 
11 =  1 +1 + 3 + 6  
\eea
$3$ is the size of a symmetric $2 \times 2 $ matrix.
$6$ is the size of a symmetric $3 \times 3$ matrix. 
More precisely the parameters form 
\bea\label{RepParameters} 
\cM = \mR^+ \times \mR^+  \times \cM_{ 2}^+ \times \cM_{ 3}^+  
\eea
where $ \mR^+ $ is the set of real numbers greater or equal to zero,  
$\cM_{ r }^+$ is the space of positive semi-definite matrices of size $r$. 
Calculating the correlators of this Gaussian model 
amounts to inverting a symmetric $2 \times 2 $ matrix, 
inverting a symmetric $3 \times 3 $ matrix, and applying 
Wick contraction rules, as in quantum field theory,  for calculating correlators. 
There is a graph basis for permutation invariant functions 
of $M$. This is explained in Appendix  B of \cite{LMT} which gives examples of 
 graph basis invariants and  representation theoretic counting formulae which make contact with 
 the sequence A052171 - directed multi-graphs with loops on any number of nodes - 
  of the Online Encyclopaedia of Integer Sequences (OEIS) \cite{OEISmulti}.

In this paper we  show how all  the linear and quadratic moments 
of the graph-basis invariants  are expressed in terms of the representation theoretic 
parameters of  (\ref{RepParameters}). We also show how some cubic and quartic graph basis invariants 
are expressed in terms of these parameters. These results are analytic expressions valid for all $D$. 

The paper is organised as follows. 
Section \ref{sec:IntroPIMT} introduces the relevant facts from the representation theory of $S_D$ 
 we need in a
fairly self-contained way, which can be read with little prior familiarity of rep theory, but only knowledge of linear algebra. This is used to define the 13-parameter family of Gaussian models 
(equations (\ref{partitionfnc}) ,(\ref{action}), (\ref{genexpvals})). 
Section \ref{sec:reptograph} calculates the expectation values of 
 linear and quadratic graph-basis invariants 
in the Gaussian model. Sections  \ref{sec:cubic} and \ref{sec:quartic} describe calculations of  expectation values of 
a selection of cubic and quartic graph-basis invariants in the model.

\section{ General permutation invariant Gaussian Matrix models }\label{sec:IntroPIMT}  

We solved a permutation invariant Gaussian 
Matrix model with $2$ linear and  $3$  quadratic parameters \cite{LMT}, obtaining analytic expressions for low order moments of permutation invariant polynomial functions of a matrix variable as a function of the 5 parameters (section 6 of \cite{LMT}).  
The linear parameters are coefficients of linear permutation invariant functions of $M$ 
and the quadratic parameters (denoted  $ \Lambda , a , b $) are coefficients of quadratic functions. We  explained the existence of a $2+11$ parameter family of models, based on the fact that there are $11$ linearly independent quadratic permutation invariant functions of a matrix. 
The  general $2+11 $-parameter family of models can be solved with the help of 
 techniques from the representation theory of $S_D$.

We give a brief informal description of the key concepts we will use here. Further information can be found in  \cite{FulHar,Hamermesh,ZeeGroup,NaimarkStern}, and we will give more precise references below.   A representation of a finite group $G$ is a pair $ ( V , D^V  ) $ consisting of a vector space $V$  and a  homomorphism $D^V$ from $G$ to the space of invertible linear operators acting on V. Physicists often speak of a representation $V$ of $G$, when the accompanying homomorphism is left implicit. 
 The homomorphism associates to each $g \in G$ a linear operator $D^V ( g )$. Distinguished among the representations of $G$ are the {\it irreducible representations} (irreps). It is known that any representation of $G$ is isomorphic to a direct sum of irreducible representations. For further explanations of these statements see Lecture 1 of \cite{FulHar}. When a representation $V$ is a direct sum of $ V_1 , V_2 , \cdots , V_k $, we express this as 
\bea
V = V_1 \oplus V_2 \oplus \cdots \oplus V_k  \, . 
\eea
This implies that the linear operators $ D^V ( g )$ corresponding to group elements $ g \in G$
can, after an appropriate choice of basis in $V$, be put in a block diagonal form where the blocks are $ D^{ V_1} ( g ) , D^{ V_2 } ( g ) , \cdots , D^{ V_k} ( g ) $. The problem of finding 
this change of basis is called ``reducing the representation $V$ into a direct sum of irreducibles''. 

Given two representations  $ ( V_1 , D^{ V_1} ) $ and  $ ( V_2 , D^{ V_2} ) $ 
of $G$, the tensor product space $V_1 \otimes V_2 $ is a representation of the product 
group  $G \times G$, which consists of pairs $ ( g_1 , g_2 ) $ with $ g_1 , g_2 \in G$. 
The product  group $ G \times G$ has a subgroup of pairs $ ( g , g )$ which is called the diagonal subgroup of $G$, denoted $ { \rm Diag } ( G ) $.  The tensor product space $ V_1 \otimes V_2 $ is also a representation of 
this diagonal subgroup (see for example Chapter 1 of \cite{NaimarkStern}). 
 The linear transformation which reduces $V_1 \otimes V_2$ into a direct sum of 
irreducibles of $ {\rm Diag } ( G ) $ is called the Clebsch-Gordan decomposition. The matrix elements of the transformation are called {\it Clebsch-Gordan coefficients}. More details on these can be found in Chapter 5 of \cite{Hamermesh}. These can be used to construct projection operators for the subspaces of 
the tensor product space corresponding to particular irreducible representations. 

In section \ref{MatNat},  we introduce the natural representation  $V_D$ of $S_D$. We note that the space
of linear combinations of the matrix variables $M_{ ij}$ is isomorphic as a vector space to 
$ V_D \otimes V_D$. We recall the known fact that 
 $V_D$ is isomorphic to a direct sum of two irreducible representations 
 \bea 
 V_D = V_0 \oplus V_H \nnm 
 \eea
and give the explicit change of basis which demonstrates this isomorphism. 
The tensor product is thus isomorphic to  a direct sum 
\bea\label{SDSDvars}  
V_D \otimes V_D = ( V_0 \otimes V_0 ) \oplus ( V_0 \otimes V_H )  \oplus ( V_H \otimes V_0 )  \oplus ( V_H \otimes V_H ) 
\eea
This leads to the definition (Equation (\ref{prodSDvars})) of $ S_D \times S_D$ covariant variables $ S^{ 00} , S_{a}^{ 0 H} ,S_{a}^{ H , 0} , S_{ab}^{ HH }$, which correspond to the four terms in the expansion (\ref{SDSDvars}). 

In section \ref{MatDiagSD},  we describe the space of linear combinations of $M_{ ij }$ as a representation
of $ {\rm { Diag } } (S_D)$: 
\bea\label{MatrixVDVD}  
&& \hbox{ Span }  \{  M_{ ij} :  1 \le  i , j  \le D \}  = V_0 \oplus V_0 \oplus V_{ H} \oplus V_H \oplus V_H \oplus V_2 \oplus V_3  \cr 
&& = \bigoplus_{ \alpha  =1}^{ 2} V_0^{ ( \alpha ) }  \bigoplus_{ \alpha  =1}^{ 3}  V_H^{ (\alpha )}  \oplus V_2 \oplus V_3 
\eea
The irreps $ V_2, V_3$ have dimensions $ ( D) ( D-3) /2 $ and $ ( D-1) (D-2) / 2$. 
The {\it multiplicity index} $\alpha $ keeps track of the fact the same irrep appears multiple times in 
the decomposition into irreducibles of $ \hbox{ Span }  ( M_{ ij} )$. The isomorphism of representations of 
$S_D$ above implies the identity relating the dimensions 
\bea 
D^2 = 2 + 3 ( D-1 ) + { D ( D-3) \over 2 } + {  (D-1 )  ( D-2 ) \over 2 } 
\eea
This decomposition leads to the definition, in equations (\ref{convV0}), (\ref{convVH}) , (\ref{convV23}),  of variables $S^{ V_i ; \alpha  } $ transforming according to 
the decomposition (\ref{MatrixVDVD}). 
 
The next key observation is to think about the vector space of quadratic polynomials in  indeterminates $ \{ x_1 , x_2 , \cdots, x_N \}$ in a way which is amenable to the methods of representation theory. 
Consider a vector space $V_N $ spanned by $ x_1 , x_2 , \cdots , x_N$. The quadratic polynomials are spanned by the set of monomials $x_i x_j$ which contains $ N ( N +1)/2$ elements. The vector space can be identified with the subspace of the tensor product $ V_N \otimes V_N$ which is invariant under the exchange of the two factors using the map 
\bea 
x_i x_j \rightarrow ( x_i \otimes x_j + x_j \otimes x_i ) \, . 
\eea
This subspace of $ V_N \otimes V_N$ is denoted by $\Sym^2 ( V_N)$. In section (\ref{MatDiagSD}) we apply this observation to the space of  quadratic polynomials in the matrix variables  $ M_{ij}$. 
They form a vector space which is isomorphic to 
$\Sym^2 ( V_D \otimes V_D )$. 

Using the decomposition 
(\ref{MatrixVDVD}), we are able to find the $S_D$ invariants by using a general theorem about invariants 
in tensor products of irreducible representations. For two irreps $V_R , V_S$, the tensor product 
 $ V_R \otimes V_S$ contains the trivial representation of 
the diagonal $S_D$ only if $ R = S$, i.e. $V_R$ is isomorphic to $V_S$, and further it is also known 
that this invariant appears in the symmetric subspace $\Sym^2 ( V_R ) \subset ( V_R \otimes V_R )  $. For further information on this useful fact, the reader is referred to Chapter 5 of \cite{Hamermesh}.

This culminates in section (\ref{QuadInvRep}) in an elegant representation theoretic 
description of the quadratic 
invariants in the matrix variables, using the linear combinations $ S^{ V_i ; \alpha }$. 
With this description in hand, we introduce a set of representation theoretic parameters for 
the 13-parameter Gaussian matrix models, see equations (\ref{partitionfnc})  and (\ref{action}). In terms of 
these parameters, the linear and quadratic expectation values of $ S^{ V_i ; \alpha }$ are simple (see 
equations (\ref{V0tad1}), (\ref{H23tad}), (\ref{QDExp1})). 
The computation of the correlators of low order polynomial invariant functions of the matrices 
then follows using Wick's theorem from quantum field theory (see for example Appendix A of \cite{Zee}).

\subsection{ Matrix variables $M_{ ij} $ and the natural  representation of $S_D \times S_D $ } \label{MatNat} 

The matrix elements $M_{ij}$, where $ i, j $ run over $ \{ 1, 2 , \cdots , D \}$
span a vector space of dimension $D^2$. It is isomorphic to the tensor product 
$ V_{ D  }  \otimes V_{ D   } $, where $V_{ D} $ is a $D$-dimensional space. 
Consider $V_{ D  } $ as a span of $D$ basis vectors $\{ e_1 , e_2 , \cdots , e_D \}$. 
This vector space $V_D$ is a  representation of $S_D$. For every permutation $ \sigma \in S_D$, there is a linear operator 
$ \rho_{ V_D} ( \sigma ) $ defined by 
\bea 
\rho_{ V_D} ( \sigma ) e_i = e_{ \sigma^{-1} ( i ) } 
\eea
on the basis vectors and extended by linearity. 
With this definition, $ \rho_{ V_D} $ is a homomorphism from $ S_D$ to linear operators acting on $ V_D$
\bea 
\rho_{ V_D } ( \sigma_1 ) \rho_{ V_D } ( \sigma_2 )  = \rho_{ V_D } ( \sigma_1 \sigma_2  )  \, . 
\eea
We introduce an  inner product $ ( . ~,~ . ) $ where the $e_i$ form an orthonormal basis
\bea\label{InnProd} 
( e_i , e_j ) = \delta_{ i j }  \, . 
\eea
We can form  the following linear combinations 
\bea 
E_0 & = &  { 1 \over \sqrt { D}  }  ( e_1 + e_2 + \cdots + e_D ) \cr 
E_{ 1} & = &  { 1 \over  \sqrt { 2 }  } ( e_1 - e_2 ) \cr 
E_{ 2} & = & { 1 \over  \sqrt{ 6 } }    ( e_1 + e_2 - 2 e_3  ) \cr 
& \vdots &  \cr 
E_{ a } & = & { 1 \over \sqrt{ a ( a+1)   } } ( e_1 + e_2 + \cdots + e_a - a e_{a+1}  )   \cr 
& \vdots & \cr 
E_{D-1}  & = & { 1 \over \sqrt{ D ( D + 1) } }  ( e_1 + e_2 + \cdots + e_{D-1}  - ( D-1)  e_D )    \, . 
\eea
 $E_0$ is invariant under the action of $S_D$ 
\bea 
\rho_{ V_D } ( \sigma ) E_0 = E_0 
\eea
since, for any $ \sigma $, we have 
\bea 
e_{ \sigma^{-1} (1) } + e_{ \sigma^{-1} (2) } + \cdots + e_{ \sigma^{-1} ( D ) } 
= e_{ 1} + e_{ 2} + \cdots + e_D \, . 
\eea
Thus the one-dimensional vector space spanned by $ E_0$ is an $S_D$ invariant vector subspace of 
$V_D$.  We can call this vector space $ V_0$. 
The vector space spanned by $ E_{ a } $, where $  1 \le  a \le ( D-1)  $, which we call $ V_H $, is also an $S_D$-invariant subspace
\bea 
\rho_{ V_D } ( \sigma  ) E_a  \in V_H \, . 
\eea
We have  a matrix $ D^H ( \sigma )  $ with matrix elements  $D^{H}_{ ab}  ( \sigma ) $  such that 
\bea\label{IntroDH}  
\rho_{ V_D } ( \sigma  ) E_a = \sum_{ b =1  }^{ D-1}  D^H_{ b a} ( \sigma )  E_b \, . 
\eea 
These matrices are obtained by using the action on the $e_i$ and the change of basis coefficients.  The vectors $ E_A $ for $  0 \le A \le D-1$ are orthonormal under the inner product (\ref{InnProd})
\bea 
( E_A , E_B ) = \delta_{ A , B }  \, . 
\eea
 All the above facts are summarised by saying that the natural representation  $ V_{ D } $ of $ S_D$ 
decomposes as an orthogonal  direct sum of irreducible representations of $ S_D$ as 
\bea 
V_{ D } = V_0 \oplus V_H \, . 
\eea
By reading off the coefficients in the expansion of the $ E_0, E_a$ in $ V_H$, we can define the coefficients 
\bea\label{defCai}  
C_{ 0 ,  i } & = &  ( E_0 , e_i )  \cr 
C_{ a  , i }  & = &  ( E_a , e_i ) 
\eea
using the inner product (\ref{InnProd}). 
They are 
\bea\label{theCs} 
C_{ 0 ,  i } & = & { 1 \over \sqrt { D } } \cr 
C_{ a , i } &=&  \cN_a  \left ( - a ~\delta_{ i , a+1}     +  \sum_{ j =1}^a \delta_{ j i }  \right ) \cr  
\cN_{a}  &= & { 1 \over \sqrt { a ( a+1) } } \, . 
\eea
 The orthonormality  means that 
\bea\label{ortho} 
\sum_{ i =1  }^{ D }  C_{ 0 , i} C_{ 0,  i } = 1 \cr 
\sum_{i=1}^{ D }  C_{ a  ,  i } C_{ b  , i } = \delta_{ a, b } \cr 
\sum_{ i =1  }^{ D }  C_{ 0, i} C_{ a, i } = 0  \, . 
\eea
The last equation implies that 
\bea\label{isumC}
\sum_{ i =1 }^{ D }  C_{ a , i } = 0  \, . 
\eea
From 
\bea 
\sum_{ A =  0}^{ D-1 } C_{ A, i } C_{ A ,  j } = 
C_{ 0 , i} C_{ 0 , i} + \sum_{ a =1}^{ D-1 }  C_{ a , i  } C_{ a ,  j} = \delta_{ i , j } 
\eea
we deduce 
\bea\label{asumCC}  
\sum_{ a  =1 }^{ D-1}  C_{ a ,  i } C_{ a ,  j } = ( \delta_{ ij} - { 1 \over D } ) \equiv F ( i , j )  \, . 
\eea
As we will see, this function $ F( i , j )$ will play an important role in calculations 
of correlators in the Gaussian model.  It is the projector in $V_{ D } $ for the subspace $V_H$, obeying 
\bea 
&& \sum_{ j =1 }^{ D }  F ( i , j ) F( j , k ) = F ( i,k )  \cr 
&& \sum_{ i =1 }^{ D }  F ( i , i ) = ( D - 1 )  \, . 
\eea

Now we will use these coefficients $C_{A , i}$ to build linear combinations of the matrix elements $M_{ i , j }$ which 
have well-defined transformation properties under $S_D \times S_D$. 
Define 
\bea\label{prodSDvars}  
S^{ 00 } & = &  \sum_{ i , j =1  }^{ D}  C_{ 0, i} C_{ 0, j} M_{ ij} = { 1 \over  D  } \sum_{ i , j = 1}^D  M_{ i j } \cr 
S^{ 0 H }_a & = & \sum_{ i , j =1  }^{ D }  C_{ 0 , i } C_{ a, j} M_{ i j } = { 1 \over \sqrt{ D } } \sum_{ i , j =1  }^{ D }  C_{ a , j } M_{ ij}  \cr 
S^{ H0  }_a & = &  \sum_{ i , j =1 }^D  C_{ a , i } C_{ 0, j } M_{ ij} = { 1 \over \sqrt{ D } } \sum_{ i , j =1 }^D  C_{ a, i} M_{ ij} \cr 
S^{ HH }_{ ab}  & = & \sum_{ i , j  =1 }^{ D }  C_{ a , i } C_{ b , j } M_{ i , j  }   \, . 
\eea
The $ a, b $ indices range over $ 1 \cdots ( D-1)$. These  variables 
are irreducible under $ S_D \times S_D $, transforming as $ V_0 \otimes V_0, V_0 \otimes V_H , V_H \otimes V_0, V_H \otimes V_H$. Under the diagonal $S_D$, the first three  transform as $ V_0, V_H , V_H$ while  
$S^{HH}_{ ab}$ form a reducible representation.

Conversely, we can write these $M$ variables in terms of the $S$ variables, using the orthogonality 
properties of the $C_{ 0 , i } , C_{ a , i} $, 
\bea 
M_{ ij} && = C_{ 0, i} C_{ 0, j} S^{ 00} + \sum_{ a =1 }^{ D-1}  C_{ 0, i } C_{ a, j} S^{ 0H}_a + \sum_{ a =1 }^{ D-1}  C_{ a , i } C_{ 0, j} S^{ H 0 }_a 
 + \sum_{ a, b =1 }^{ D -1 } C_{ a, i } C_{ b, j} S^{HH}_{ ab}    \cr 
 && = { 1 \over D } S^{ 00} + { 1 \over \sqrt { D } } \sum_{ a =1 }^{ D-1}  C_{ a, j} S^{ 0H}_a
 + { 1 \over \sqrt{ D }} \sum_{ a =1 }^{ D-1}  C_{ a , i }  S^{ H 0 }_a  
 + \sum_{ a, b =1 }^{ D -1 } C_{ a, i } C_{ b, j} S^{HH}_{ ab}   \, . 
\eea

The next step is to consider quadratic products of these $S$-variables, and identify the products which are invariant. 
In order to do this we need to understand the transformation properties of the above $S$ variables 
in terms of the diagonal action of $S_D$. It is easy to see that $ S^{ 00}$ is invariant. $ S^{0H}_{  a } $ and $ S^{H0}_{ a } $ both have a single $ a $ index
running over $ \{ 1 , 2 , \cdots , ( D-1) \} $, and they transform in the same way as $V_H$. 
The vector space spanned by $S^{ HH}_{ ab } $  form a space of dimension $ ( D-1)^2 $ which is 
\bea 
V_{ H } \otimes V_{ H  }  \, . 
\eea
Permutations act on this as 
\bea 
\sigma ( S^{HH}_{ ab } ) = \sum_{ a_1 , b_1 } D^H_{a_1 a }  ( \sigma ) ~ D^H_{b_1 b } ( \sigma  ) ~  S^{HH}_{ a_1 , b_1} \, . 
\eea 
using the matrices $D^{H} ( \sigma )$ introduced in (\ref{IntroDH}). 

\subsection{  Decomposition of matrix variables as irreducible representations of $ { \rm Diag }  ( S_D ) \subset S_D \times S_D $  } \label{MatDiagSD} 

In this section we will perform a further change of variables to introduce variables $ S^{V_i ; \alpha } $ which transform according to irreps of the diagonal subgroup  $ { \rm Diag }  ( S_D ) \subset S_D \times S_D $. 

\begin{itemize} 

\item  The representation space  $ V_H \otimes V_H $ can be decomposed into irreducible representations (irreps)  of the diagonal $S_D$ action as 
\bea\label{decompsq}  
V_{ H  } \otimes V_{ H } = V_0 \oplus  V_H \oplus  V_2 \oplus V_{ 3} \, . 
\eea
In Young diagram notation for irreps of $S_D$, listing the row lengths of the Young diagram, we have 
\bea 
&& V_0 \rightarrow [ D ] \cr 
&& V_H \rightarrow [ D-1, 1 ]  \cr 
&& V_{ 2 } \rightarrow [ D-2,2] \cr 
&& V_{ 3} \rightarrow [ D-2,1,1]  \, . 
\eea

\item These irreps are known to have dimensions $ 1 , ( D-1) ,  { D ( D-3) \over 2 } , { ( D-1) ( D-2) \over 2 }  $. 
They add up to $ ( D-1)^2$ which is the dimension of $ V_H^{ \otimes 2 } $. 

\item The vector $ \sum_{ a =1}^{ D-1} E_{ a }  \otimes E_{ a} $ is  invariant under  the diagonal action of $ \sigma $ on $ V_H \otimes V_H$.  Using the fact that $V_H$ is a subspace of $V_D$ described 
by the coefficients $C_{ a , i }$ defined in (\ref{defCai}),  the action of $ \sigma $ on $ V_H$ is given by 
\bea 
D^{ H}_{ ab} ( \sigma ) = ( E_{ a} , \sigma E_b ) = 
\sum_{ i  =1 }^{ D }  C_{ a , i } C_{ b , \sigma( i ) }  \, . 
\eea
These can be verified to satisfy the homomorphism property 
\bea 
\sum_{ b =1 }^{ D-1} D^{ H}_{ ab} ( \sigma ) D^{H}_{ b c } ( \tau ) = D^{ H}_{ ac} ( \sigma \tau )  \, . 
\eea
We also have $ D^{H}_{ ab }  ( \sigma^{-1}  )  = D^H_{ ba}  ( \sigma ) $. 
Using these properties, we can show that $  \sum_{ a} E_{ a } \otimes E_a $ is invariant under the diagonal action. 
The vector 
\bea\label{HH0}  
 { 1 \over  \sqrt { D -1 }   } \sum_{ a =1  }^{ D-1}  E_a \otimes E_a  
\eea
has unit norm, using the inner product on $ V_D \otimes V_D$ obtained from (\ref{InnProd}), and defines 
a normalized vector in the $V_0$ subspace of the direct sum decompsotion of $ V_H \otimes V_H$
given in (\ref{decompsq}). From this expression, we can read off the Clebsch-Gordan coefficients 
for  the trivial representation $ V_0 $ in $ V_H \otimes V_H$
\bea 
C_{~ a , ~ b }^{ H,H \rightarrow V_0  }  = { \delta_{ ab } \over \sqrt{ D -1 } }  \, . 
\eea
Using these we define $ S^{ H  H \rightarrow V_0 }$ as 
\bea\label{SHH0}  
&& S^{ H  H \rightarrow V_0 } = \sum_{ a , b =1}^{ D-1}   C_{ ~ a , ~ b }^{ ~H, H \rightarrow V_0  } S_{ ab }^{ HH }  \cr 
&&  = { 1 \over \sqrt {D-1} } \sum_{ a =1 }^{ D-1}  S^{ H H }_{ a a } 
\eea

\item The vectors in the  $ V_H$ subspace on the RHS of the direct sum decomposition  (\ref{decompsq}) are some linear combinations 
\bea\label{HHH}  
\sum_{ b , c  =1 }^{ D-1}  C^{ H,H \rightarrow H  }_{ ~b ,~c  ~;~a } S^{HH}_{ b  c} \equiv S^{ H,H \rightarrow H }_{  a }  \, . 
\eea
The coefficients $ C^{ H, H \rightarrow H    }_{ ~b, ~c ~;~  a}  $ are some representation theoretic numbers (called Clebsch-Gordan coefficients ) 
which satisfy the orthonormality condition 
\bea 
\sum_{ b , c =1 }^{ D-1}   C^{ H, H \rightarrow H  }_{ ~b, ~c ~;~   a  } C^{ H,H \rightarrow H }_{ ~b, ~c ~;~    d } = \delta_{ a  , d}  \, . 
\eea 

As shown in Appendix \ref{repVH}, these Clebsch-Gordan coefficients are proportional to  
$C_{ a ,  b ,  c } \equiv \sum_{ i } C_{  a  ,   i } C_{ b ,   i } C_{ c ,   i }$
\bea 
C^{HH \rightarrow H  }_{ ~a , ~b ~; ~ c } = \sqrt { D \over ( D -2 ) }  C_{ a,  b ,  c } \, . 
\eea
It is a useful fact  that the Clebsch-Gordan coefficients for $V_H \otimes V_H \rightarrow V_H$ can be 
usefully written in terms of the  $C_{ a, i}$ describing $V_H$ as a subspace of the natural 
representation. This has recently played a role in the explicit description of 
a ring structure on primary fields of free scalar 
 conformal field theory \cite{FFPR}. It would be interesting to 
explore the more general construction of  explicit Clebsch-Gordan coefficients and projectors in the 
representation theory of $S_D$ in terms of the $C_{ a, i}$.

\item Similarly for $ V_2, V_3 $ we have corresponding vectors and Clebsch-Gordan coefficients 
\bea\label{HHV2} 
\sum_{ b , c  =1 }^{ D -1  }  C^{ H , H \rightarrow  V_2  }_{ ~b,~ c  ~;~  a  } S_{ b, c} \equiv S^{ HH\rightarrow V_2 }_{  a }  
\eea
where $ a $ ranges from $ 1 $ to $ \Dim ( V_2 ) = {  D( D-3) \over 2 } $. 
We have the orthogonality property 
\bea 
\sum_{ b , c =1  }^{ D -1  }  C^{ H, H \rightarrow V_2 }_{ ~b , ~c ~;~  a_1  }C^{ H , H \rightarrow V_2  }_{ ~b , ~c ~;~  a_2  } = \delta_{ a_1 , a_2 }  \, . 
\eea
And for $ V_3$
\bea\label{HHV3}  
&& \sum_{ b , c =1  }^{ D -1  }  C^{ H , H \rightarrow  V_3 }_{ ~b, ~c  ~;~  a  } S^{HH}_{ bc} \equiv S^{ HH \rightarrow V_3  }_a  \cr 
&& \sum_{ b , c =1 }^{ D  -1 }  C^{ H , H \rightarrow V_3 }_{ ~b, ~c ~;~  a_1  }C^{ H , H \rightarrow V_3 }_{ ~b, ~c  ~;~ a_2 } = \delta_{ a_1 , a_2  } \, . 
\eea
Here the $a, a_1 , a_2 $ runs over $ 1$ to $   { ( D-1 ) ( D-2 )  \over 2 } $.

\item The projector for the subspace of $ V_H \otimes V_H$ transforming as $V_H$ under the diagonal $S_D$ is 
\bea 
( P^{H,H \rightarrow H} )_{ a, b ;  c, d} && = \sum_{ e =1  }^{D-1}   C^{ H , H \rightarrow H}_{  ~a , ~b ~;~ e   }   C^{ H , H \rightarrow H}_{  ~c , ~d ~;~ e   }   \cr 
&& = { D \over ( D-2) }  \sum_{ e =1 }^{ D -1}  C_{  a , b ,  e   }   C_{  c , d , e   } \, . 
\eea

\item The projector  $P^{ H, H \rightarrow  V_0} $ for $ V_0$ in $V_H \otimes V_H$ is 
\bea
( P^{ H, H \rightarrow  V_0} )_{a, b ; c , d}  = { 1 \over ( D-1) } \delta_{ a, b} \delta_{ c, d} \, . 
\eea
 $V_{ [ n-2,1,1] } = V_3 $ is just the anti-symmetric of $ V_{ H} \otimes V_H$. It is the orthogonal complement to 
 $ V_H \oplus V_0 $ inside the symmetric subspace of $V_H \otimes V_H$ which is invariant 
 under the swop of the two factors (often denoted $ \Sym^2 ( V_H) $) 
 \bea 
&& (  P^{ H , H \rightarrow  V_2} ) = 
( 1 - P^{H , H \rightarrow H } - P^{H , H \rightarrow V_0} ) { ( 1 + s )\over 2 }  \, 
\eea
where the swop $s $ acting on $ V_{ H} \otimes V_H$ has matrix elements 
\bea
   ( s)_{a , b ; c, d} = \delta_{a,  d }  \delta_{b , c }  \, . 
 \eea 
The quadratic invariant corresponding to $V_2$ is 
\bea 
S^{HH}_{ab} ( P^{H , H \rightarrow V_2} )_{a, b ; c , d} S^{HH}_{ cd} \, . 
\eea
The quadratic invariant corresponding to $ V_3$ is similar. 
We just have to calculate 
\bea 
P^{ H , H \rightarrow  V_3} = { 1 \over 2 } ( 1 - s ) \, . 
\eea

\item The inner product 
\bea 
\langle  M_{ ij} , M_{ kl} \rangle  = \delta_{ ik } \delta_{ j l } 
\eea
is invariant under the action $ \sigma ( M_{ i , j } ) = M_{ \sigma^{-1} ( i ) , \sigma^{-1} ( j ) } $.
\bea 
\langle  \sigma ( M_{ ij} ) , \sigma ( M_{ kl} ) \rangle = \langle  M_{ ij} , M_{ kl} \rangle \, .  
\eea

\item The following is an important fact about invariants. Every irreducible representation of $ S_D$, let us denote it by $ V_{ R } $ 
 has the property  that 
 \bea 
 \Sym^2 ( V_{ R }  ) 
 \eea
contains the trivial irrep once. This invariant is  formed by taking the sum 
over an orthonormal basis $\sum_{ A }  e^V_{ A } \otimes  e^V_{ A } $. The invariance is proved as follows
\bea
 D^{ V \otimes V } ( \sigma ) \sum_A  e^V_{ A} \otimes e^V_A 
& =&  \sum_A D^V ( \sigma ) e^V_A \otimes D^V ( \sigma ) e^V_A \cr 
&  = &  \sum_{ A } \sum_{ B , C } D^V_{BA } ( \sigma ) D^V_{ CA} ( \sigma ) e^V_B \otimes e^V_{ C} \cr 
& =  & \sum_{ A , B ,  C  } D^V_{BA } ( \sigma ) D^V_{ AC} ( \sigma^{-1}  ) e^V_B \otimes e^V_{ C}\cr 
&  =  &   \sum_{ B , C } \delta_{ B , C }   e^V_{ B } \otimes e^V_C  \cr 
& = & \sum_A  e^V_{ A} \otimes e^V_A \, . 
\eea
In the first equality we have used the definition of the diagonal action of $ \sigma $ on the tensor product space.

\item To summarize the matrix variables $M_{ ij}$ can be linearly transformed to 
the following variables, organised according to representations of  the diagonal $S_D$
\bea\label{IrrepsFirst}  
\hbox{ Trivial rep:}  ~~& S^{ 00  }  , S^{H H \rightarrow V_0  } \cr 
\hbox{ Hook rep:} ~~ & S^{0 H  }_{ a} , S^{ H0}_{  a  }, S^{ HH \rightarrow  H }_{ a }\cr  
\hbox{ The rep $V_2$:} ~~ & S^{HH\rightarrow V_2}_a  \cr 
\hbox{ The rep $V_3$:} ~~ & S^{ HH \rightarrow V_3}_a  \, . 
\eea

\item For convenience, we will also use simpler names 
\bea\label{convV0}  
&&  S^{ V_0  ;  1  } = S^{00 }   \cr 
&& S^{ V_0  ;  2  } = S^{H H \rightarrow V_0 }  
\eea
where we introduced labels $1,2$ to distinguish two occurrences of the trivial irrep $V_0$
in the space spanned by the $M_{ij}$.  The variables $ S^{ 0 , 0 } , S^{ H , H \rightarrow V_0 } $ 
were first introduced in (\ref{prodSDvars}) and (\ref{SHH0}) respectively. 
We will also use 
\bea\label{convVH} 
&&  S^{  H; 1  }_a = S^{0, H \rightarrow H}_{   a}   \equiv  S^{0 H }_{   a}   \cr
&& S^{  H;2  }_a =  S^{ H,0 \rightarrow H }_{a  } \equiv S^{ H  0  }_{   a}   \cr 
&&  S^{  H;3  }_a = S^{ H,H \rightarrow  H }_{ a }
\eea
where we introduced labels $ 1,2,3$ to distinguish the three occurrences of $V_H$ 
in  the space spanned by $M_{ ij}$.  The variables $ S^{ 0 H }_{ a} , S^{ H  0 }_a $ were 
introduced earlier in (\ref{prodSDvars}). 
For the multiplicity-free cases, we introduce 
\bea\label{convV23} 
&&   S^{ V_2}_a =  S^{HH\rightarrow V_2}_a  \cr 
&&   S^{ V_3 }_a = S^{HH\rightarrow V_3}_a  \, . 
\eea

The $ M_{ ij} $ variables can be written as linear combinations of 
the $S$ variables. 
Rep-basis expansion of $M_{ij}$ is 
\bea\label{MtoS} 
M_{ ij} && = C_{  0 ,  i  } C_{ 0 ,  j } S^{ 00}  + \sum_{ a , b =1}^{ D-1} C_{ a  ,  i } C_{  b ,  j  } S^{HH}_{ ab } 
+ \sum_{ a=1}^{ D-1}  C_{0  , i} C_{ a  , j } S^{0H}_{a}  + \sum_{ a=1}^{ D-1}  C_{ a , i } C_{ 0 , j }  S^{H0}_{ a} \cr 
 && = { 1 \over D } S^{ 00} + { 1 \over \sqrt { D } } \sum_{ a =1 }^{ D-1}  C_{ a , j} S^{ 0H}_a
 + { 1 \over \sqrt{ D }} \sum_{ a =1 }^{ D-1}  C_{ a,  i }  S^{ H 0 }_a  
 + \sum_{ a, b =1 }^{ D -1 } C_{ a , i } C_{ b , j} S^{HH}_{ ab}  \cr 
  && = { 1 \over D } S^{ 00} + { 1 \over \sqrt { D } } \sum_{ a =1 }^{ D-1}  C_{ a, j} S^{ 0H}_a
 + { 1 \over \sqrt{ D }} \sum_{ a =1 }^{ D-1}  C_{ a ,  i }  S^{ H 0 }_a    \cr 
 && \hskip2cm 
 +  \sum_{ a , b   } C_{  a , i  } C_{    b ,  j  } 
  \sum_{ V \in \{ V_0 , V_H , V_2 , V_3 \} } \sum_{ c =1  }^{ \Dim V }   C_{ ~a, ~b ~;~ c  }^{ H H \rightarrow V }~ S^{HH \rightarrow V }_{ c}  \cr  
 && = { 1 \over D } S^{ 00} + { 1 \over \sqrt { D } } \sum_{ a =1 }^{ D-1}  C_{ a , j} S^{ 0H}_a
 + { 1 \over \sqrt{ D }} \sum_{ a =1 }^{ D-1}  C_{ a , i }  S^{ H 0 }_a   + { 1 \over \sqrt { D-1  }  }  \sum_{ a =1}^{ D-1}  C_{  a , i  } C_{   a , j } S^{HH \rightarrow V_0 } \cr 
 && 
 + \sum_{ a, b =1 }^{ D -1 } C_{ a ,  i } C_{ b,   j }  \sum_{ c  =1 }^{ D-1 }     C_{ ~a, ~b ~;~ c }^{ HH \rightarrow H }
  S^{ HH \rightarrow H}_{ c} + 
 \sum_{ a, b =1 }^{ D -1 } C_{ a  ,  i } C_{ b  , j }   \sum_{ c=1}^{ \Dim   V_2}  C_{ ~a , ~b ~;~ c }^{ HH \rightarrow V_2 }
  S^{ HH \rightarrow V_2 }_{ c }  \cr 
  && \hskip2cm   ~ + \sum_{ a, b =1 }^{ D -1 } \sum_{ c = 1 }^{ \Dim V_3 }   C_{ a  ,  i } C_{ b ,  j }    C_{ ~a , ~b ~;~ c }^{ HH \rightarrow V_3 }
  S^{ HH \rightarrow V_3 }_{ c} \, . \cr 
  && 
\eea
In going from first to second line, we have used the fact that the transition from the natural 
representation to the trivial representation is given by simple constant coefficients
\bea 
C_{ 0 , j } = { 1 \over \sqrt{ D }  } \, . 
\eea
In the third line, we have used the Clebsch-Gordan coefficients for $ V_H \otimes V_H \rightarrow V$, 
obeying the orthogonality 
\bea
\sum_{ a, b =1}^{ D-1}   C^{ H, H \rightarrow V}_{ ab c } C^{ H , H \rightarrow V}_{ ab c' } = \delta_{ c c'} \, . 
\eea
For $V = V_0$, which is one dimensional, we just have
\bea 
C_{ ab }^{ HH \rightarrow V_0  }  = { \delta_{ ab } \over \sqrt{ D -1 } } \, , 
\eea
in accordance with (\ref{HH0}).
The index $c$ ranges over a set of orthonormal basis vectors for  the irrep $V$, i.e. extends over a range equal to the dimension of $V$, denoted $ \Dim V$. 
It is now useful to collect together the terms corresponding to each irrep $V_0 , V_H , V_2 , V_3 $
\bea 
M_{ ij} && = \left ( { 1 \over D } S^{ 00} + { 1 \over \sqrt { D-1 }  }  \sum_{ a =1 }^{ D-1}  C_{ a ,  i  } C_{  a ,  j } S^{HH \rightarrow 0 } \right ) \cr 
&&  + \left (   { 1 \over \sqrt { D } } \sum_{ a =1 }^{ D-1}  C_{ a, j} S^{ 0H}_a
 + { 1 \over \sqrt{ D }} \sum_{ a =1 }^{ D-1}  C_{ a ,  i }  S^{ H 0 }_a  + 
  \sum_{ a, b , c =1 }^{ D -1 } C_{ a ,   i } C_{ b  ,  j }    C_{ ~a, ~b ~;~ c }^{ HH \rightarrow H }
  S^{ HH \rightarrow H}_{ c} \right ) \cr 
  && +  \sum_{ a, b =1 }^{ D -1 } \sum_{ c =1}^{ \Dim V_2 }  C_{ a ,  i } C_{ b ,  j }    C_{ ~a, ~b ~;~ c }^{ HH \rightarrow V_2 }
  S^{ HH \rightarrow V_2 }_{ c } +  \sum_{ a, b =1 }^{ D -1 } \sum_{ c=1}^{ \Dim V_3} C_{ a  ,  i } C_{ b  ,  j }    C_{ ~a , ~b  ~;~ c }^{ HH \rightarrow V_3 }
  S^{ HH \rightarrow V_3 }_{ c} \, .   \cr 
  && 
\eea
Using the notation of (\ref{convV0}), (\ref{convVH}), (\ref{convV23}) , we write this 
as 
\bea\label{expMS}  
&& M_{ ij } = \left ( { 1 \over D } S^{ V_0 ; 1  }  + { 1 \over \sqrt { D -1  }  }  \sum_{ a=1 }^{ D-1}  C_{  a ,  i  } C_{   a  ,  j } S^{  V_0 ; 2  } \right ) \cr 
&&  + \left (   { 1 \over \sqrt { D } } \sum_{ a =1 }^{ D-1}  C_{ a , j} S^{  H ; 1  }_a
 + { 1 \over \sqrt{ D } } \sum_{ a =1 }^{ D-1}  C_{ a ,  i }  S^{ H ; 2   }_a  + 
  \sum_{ a,  b , c =1 }^{ D -1 } C_{ a ,  i } C_{ b ,  j }    C_{ ~a, ~b  ~;~ c }^{ HH \rightarrow H }
  S^{  H ; 3  }_{ c} \right ) \cr 
  && +  \sum_{ a, b =1 }^{ D -1 } \sum_{ c=1}^{ \Dim V_2 } C_{ a ,  i } C_{ b  , j }    C_{ ~a, ~b ~; ~ c }^{ HH \rightarrow V_2 }
  S^{ HH \rightarrow V_2 }_{ c } +  \sum_{ a, b =1 }^{ D -1 } \sum_{ c=1}^{ \Dim V_3 }  C_{ a  ,  i } C_{ b ,  j } 
     C_{ ~a, ~b ~;~ c }^{ HH \rightarrow V_3 }
  S^{ HH \rightarrow V_3 }_{ c}  \, .  \cr 
&& 
\eea
 
\item The  discussion so far  has included explicit bases for $V_H$ inside 
$V_{ D } $ which are easy to write down. A key object in the above discussion 
is the projector $ F ( i , j ) $ defined in (\ref{asumCC}).  
 For the irreps $V_2, V_3$ which appear in $V_H \otimes V_H$, 
  we will not need to  write down explicit bases. Although 
   Clebsch-Gordan coefficients for $H , H \rightarrow V_2$ 
   and $H , H \rightarrow V_3$ appear in some of the  above formulae, we will only need some of their orthogonality properties rather than their explicit forms. The projectors for $V_2, V_3$ in $V_D \otimes V_D$ can be written in terms of the $F( i , j )$, and it is these projectors which play a role 
   in the correlators we will be calculating.

\end{itemize} 

\subsection{Representation theoretic description of quadratic invariants }\label{QuadInvRep} 

With the above background of facts from representation theory at hand, we can give a 
useful description of quadratic invariants. 
Quadratic invariant functions of $M_{ ij}$ form the invariant subspace of  $ \Sym^2 ( V_{ D }  \otimes  V_{ D } )   $  since $ M_{ ij}$ transform as
$V_D \otimes V_D$. 
\bea 
&& ( V_{ D } \otimes V_{ D  } ) = ( V_0 \oplus V_H ) \otimes ( V_0 \oplus V_H ) \cr 
&& = ( V_{00} \oplus V_{ 0 , H }^H \oplus V_{ H , 0 }^{ H } \oplus V_{ H , H }^{ 0 } \oplus 
V_{ H , H }^{ H } \oplus V_{ H , H }^{ V_2 } \oplus V_{H, H  }^{ V_3 }   ) \, . 
\eea
So there are two copies of $V_0$, namely $V_{ 0 , 0 }^0 , V_{ H , H}^0 $.
$ \Sym^2 ( V_{0,0}^0  \oplus V_{H,H}^0  ) $ contains three invariants : 
\bea 
 ( S^{ 0  0 } )^2  & = &  ( S^{  V_0 ; 1  } )^2\cr 
 ( S^{ 0 0 } S^{ HH \rightarrow 0 } ) &  = &  S^{  V_0 ; 2  }S^{ V_0 ; 1  } =
 S^{  V_0 ; 1  }S^{  V_0 ; 2  }
 \cr  
 ( S^{ HH \rightarrow 0 } )^2 &  =  & ( S^{  V_0 ; 2  } )^2 \, . 
\eea
These are all easy to write in terms of the original matrix variables, 
using the formulae for $S$-variables in terms of $M$ given earlier. The relevant equations are 
(\ref{convV0}),(\ref{convVH}),(\ref{convV23}) where $S$-variables for irreps of  $ { \rm Diag }  ( S_D  ) \subset S_D \times S_D $ along with multiplicity labels are introduced, and earlier equations 
(\ref{prodSDvars}), (\ref{HH0}) , (\ref{HHH}), (\ref{HHV2}) which introduce $S$-variables labelled 
by $ S_D \times S_D$ irreps. These latter are more directly related to the Matrix variables, but the 
former are needed to give an elegant description of the quadratic invariants.

The general invariant quadratic function of the $S^{ V_0 ; \alpha }$ variables is 
\bea 
\sum_{ \alpha , \beta =1 }^{ 2 } ( \Lambda_{ V_0} )_{ \alpha \beta } S^{ V_0 ; \alpha } S^{ V_0 ; \beta } 
\, . 
\eea
$ \Lambda_{V_0} $ is a $ 2 \times 2 $  symmetric  matrix. As we will see later, in defining the Gaussian model, this matrix will be restricted to be positive semi-definite.

There are three copies of $V_{ H} $, namely $  V_{ 0 , H }^H  , V_{ H , 0 }^{ H } , V_{ H , H }^H $.
These lead to $6$ invariants: 
\bea
 \sum_{ a } S^{ 0 , H \rightarrow H}_{ a }   S^{ 0 , H \rightarrow H}_{ a }
 & = &  \sum_{ a} S^{  H ;1  }_a S^{  H ;1  }_a   \cr 
  \sum_{ a }  S^{  H  , 0 \rightarrow H}_{ a }   S^{ H , 0  \rightarrow H}_{ a }  & = &  
\sum_{ a} S^{  H ;2  }_a S^{  H ;2  }_a  \cr 
\sum_a S^{ H, H \rightarrow H}_{ a } S^{ H, H \rightarrow H}_{ a } 
& = &   \sum_{ a} S^{ H ;3  }_a S^{ H ;3  }_a   \cr 
\sum_{ a } S^{ 0 , H \rightarrow H }_{ a } S^{H, 0 \rightarrow H  } _{ a  }
& = &  \sum_{ a} S^{  H ;1  }_a S^{  H ;2 }_a ~~ = ~~ \sum_{ a} S^{  H ;2  }_a S^{  H ;1  }_a 
 \cr 
\sum_{ a } S^{ 0 , H \rightarrow H }_a  S^{ H , H \rightarrow H  }_a   & = & 
\sum_{ a} S^{  H ;1  }_a S^{  H ;3 }_a  ~~ = ~~ \sum_{ a} S^{ H ;3 }_a S^{  H ;1  }_a 
 \cr  
\sum_{ a  } S^{ H , 0 \rightarrow H_{ a } } S^{ H , H \rightarrow H  }_a
& = & \sum_{ a} S^{  H ;2  }_a S^{  H ;3  }_a ~~ = ~~ \sum_{ a} S^{  H ;3 }_a S^{  H ;2  }_a \, . 
\eea 
The sum over $a$ runs over $(D - 1) = \Dim ( V_H ) $ elements of a basis for $V_H$. 
Thus the general  quadratic invariants arising from the $H$ representation among the $M_{ ij}$  are  
\bea 
\sum_{ \alpha , \beta =1  }^3  (\Lambda_{ H })_{ \alpha \beta } \sum_a S^{  H ; \alpha  }_a S^{  H ; \beta  }_a  \, . 
\eea
We introduced parameters $ (\Lambda_{ H })_{ \alpha \beta }  $ forming a symmetric $ 3 \times 3 $ matrix. 
When we define the general  Gaussian measure, we will see that this matrix will be required to be a positive definite matrix. 

The quadratic invariants constructed from the $V_2,V_3$ variables are 
\bea 
&& ( \Lambda_{ V_2} )  \sum_{ a =1 }^{ \Dim V_2 }  S^{ V_2 }_{ a  }  S^{ V_2 }_{ a  }  \cr 
&& ( \Lambda_{ V_3} ) \sum_{ a =1 }^{ \Dim V_3 } S^{ V_3 }_{ a  }  S^{ V_3 }_{ a  } \, . 
\eea
When we define the  general Gaussian measure, we will take the parameters $ \Lambda_{ V_2} , \Lambda_{ V_3} $ to obey $ \Lambda_{ V_2} , \Lambda_{ V_3} \ge 0$.

\subsection{ Definition of the Gaussian models }\label{DefGauss}

The measure $dM$  for integration over the matrix variables $M_{ij}$ is taken to be the Euclidean measure on $ \mR^{ D^2}$ parametrised by  the $D^2$ variables 
\bea 
dM \equiv \prod_{ i } dM_{ ii} \prod_{ i \ne j } dM_{ ij} \, . 
\eea
Since the variables $S^{V ; \alpha }_{ a }$ defined in 
 (\ref{convV0}), (\ref{convVH}), (\ref{convV23}) are given by an orthogonal change of basis, we can show that  
\bea\label{dSmeas1}  
dM = dS^{V_0; 1 }  dS^{V_0 ; 2 }  \prod_{ a = 1}^{ \Dim V_H }  dS^{ H ;1}_{a} dS^{H;2}_{a} dS_a^{H ; 3 }  \prod_{ a =1 }^{ \Dim V_2}  dS_{a}^{V_2  } \prod_{ a =1  }^{ \Dim V_3 }  dS_{ a }^{ V_3 }   \, . 
\eea
Indeed writing $ M_A$ for the matrix variables, where $A$ runs over the $D^2$ pairs $(i, j )$
and $ S_{B} $ for $S_D$-covariant variables, where $B$ runs over all the factors in (\ref{dSmeas1}),  
we have 
\bea 
dM = \prod_A dM_{A} = | \det J |  \prod_{B }  dS_B 
\eea
with 
\bea 
J_{ AB } =  { \partial M_A \over \partial S_B  } \, . 
\eea
Now the $S_B$  variables are obtained from $M_A$ by an orthogonal basis 
change, and symmetric group properties also allow the matrix to be chosen to be real. 
This implies that the matrix is orthogonal
\bea 
JJ^{T} = 1 \, . 
\eea
Hence $ \det J $  has magnitude $1$, and we have the claimed identity (\ref{dSmeas1}).

 The model is defined by integration. The partition function  is 
\bea\label{partitionfnc}  
\cZ ( \mu_{ 1} , \mu_2 ; \Lambda_{V_0} , \Lambda_{ H} , \Lambda_{ V_2} , \Lambda_{ V_3} )  = 
\int dM e^{ - \cS } 
\eea
 where the action is a combination of linear and quadratic functions.
\bea\label{action}  
&& \cS  =  - \sum_{ \alpha =1  }^2  \mu^{ V_0}_{ \alpha } S^{ V_0 ; \alpha } 
         +  { 1 \over  2 }  \sum_{ \alpha , \beta = 1 }^{ 2 } S^{ V_0 ;  \alpha } ( \Lambda_{V_0})_{ \ls[ \alpha \beta] } S^{ V_0 ;  \beta  } + { 1 \over 2 } \sum_{ a = 1 }^{ D-1  } \sum_{  \alpha , \beta = 1 }^{ 3 } S^{ H ;  \alpha }_{ a  } (\Lambda_H)_{ \ls[\alpha \beta]  } S^{ H ;  \beta }_{ a   } \cr 
         && 
          + { 1 \over 2 } \Lambda_{ V_2} \sum_{ a =1 }^{  ( D-1) ( D-2) /2 }  S^{ V_2 }_{ a }  S^{ V_2 }_{ a }             +  { 1 \over 2 }  \Lambda_{ V_3 } \sum_{ a =1 }^{ D (D-3)/2 }   S^{ V_3 }_{ a }  S^{ V_3}_{   a } \, . 
\eea

The expectation  values of permutation invariant polynomials $ f ( M ) $ are defined 
by 
\bea\label{genexpvals}  
\langle f (M ) \rangle = { 1 \over \cZ } \int dM e^{ - \cS }   f ( M)  \, . 
\eea
These expectation values can be computed using standard techniques from quantum field theory, specialised to matrix fields in zero space-time dimensions (See Appendix \ref{app:Wick} for some explanations).  Textbook discussions of these techniques are given, for example in  \cite{Peskin},\cite{Zee}. 
For linear functions, the  non-vanishing expectation values are  those of the invariant variables, which transform as $ V_0$ under the $S_D$ action
\bea\label{V0tad} 
\langle S^{ V_0 ; \alpha } \rangle = \sum_{ \beta } ( \Lambda^{-1} )_{  \alpha \beta  } ~  \mu_{ \beta } \, . 
\eea
We introduce the definition 
\bea\label{V0tad1} 
\tilde \mu_{ \alpha } \equiv  \sum_{ \beta } ( \Lambda^{-1} )_{  \alpha \beta  } ~  \mu_{ \beta } \,  .
\eea
We have defined variables $ \wmuo , \wmut$ for convenience. 
The variables transforming according to $V_H , V_2, V_3 $ have vanishing expectation values
\bea\label{H23tad} 
&& \langle S^{ H ; \alpha }_{  a } \rangle = 0 \cr  
&& \langle S^{ V_2 }_{ a } \rangle = 0 \cr 
&& \langle S^{ V_3  }_{ a } \rangle = 0 \, . 
\eea
The  quadratic expectation values  are 
\bea\label{QDExp1}   
\langle S^{ V_i ; \alpha } S^{ V_j ; \beta } \rangle = 
\langle S^{ V_i ; \alpha } S^{ V_j ; \beta } \rangle_{ \con} + 
\langle  S^{ V_i ; \alpha } \rangle  ~ \langle S^{ V_j ; \beta } \rangle  \, . 
\eea
where\label{QDExp2}  
\bea 
&& \langle S^{ V_i ; \alpha}_{ a }S^{ V_j ; \beta  }_{ b } \rangle_{ \con}  = \delta ( V_i , V_j ) 
  (\Lambda_{V_i}^{-1})_{ \alpha \beta }   \delta_{ a b } \, . 
\eea
The notation $ \langle . . \rangle_{ \con} $ is explained in the Appendix \ref{app:Wick}. 
The $ V_i , V_j  $ can be $ V_0  , V_H , V_2  , V_3 $. The delta function means that 
these expectation values vanish unless the two irreps $V_i, V_j $ are equal.  While $ \delta_{ a b  } $ is the identity in the state space for each $V_i$. The fact that the mixing matrix in the multiplicity 
indices $ \alpha , \beta $ is the inverse of the coupling matrix  $ \Lambda_V $ is a special (zero-dimensional) case of a standard result in quantum field theory, where the propagator is the inverse of the operator defining the quadratic  terms in the action.  The decoupling between different irreps 
follows because of the factorised form of the measure $ dM e^{ - \cS } $ in (\ref{partitionfnc}).

 The requirement of an $S_D$ invariant Gaussian measure has led us to define variables $S^{ V , \alpha }$, 
 transforming in irreducible representations of $S_D$. The action is simple in terms of these variables. 
 This is reflected in the fact that the above  one and two-point functions are simple 
 in terms of the parameters of the model. 

When $ \Lambda_{ V_2} > 0 , \Lambda_{ V_3 } >  0 $ and $ \Lambda_{ H } , \Lambda_{ V_0 }  $ are positive-definite  real symmetric matrices (i.e  real symmetric matrices with positive eigenvalues), then the partition function $ \cZ $ is well defined as well as the numerators in the definition of $ \langle f ( M ) \rangle $.  We can relax these conditions, allowing $ \Lambda_{ V_2 } , \Lambda_{ V_3 } \ge 0 $ 
and $ \Lambda_{ H } , \Lambda_{ V_0 } $ positive semi-definite, by appropriately restricting the $f(M)$ we consider. For example, if $ \Lambda_{ V_2 } =0 $, we consider functions $ f ( M ) $ which do not depend on $S^{ V_2 }$, which ensures that the ratios defining $ \langle f ( M ) \rangle $ are well-defined. 
 
Thus the complete set of constraints on the representation theoretic parameters are 
\bea 
&& \rm { Det } ( \Lambda_{ V_0 } ) \ge 0  \cr 
&& \rm { Det } ( \Lambda_{ V_H } ) \ge 0 \cr 
&& \Lambda_{ V_2 } \ge 0 \cr 
&& \Lambda_{ V_3} \ge 0  \, . 
\eea
More explicitly 
\bea 
&& ( \Lambda_{ V_0} )_{ \ls[11] } ( \Lambda_{ V_0} )_{ \ls[ 22 ] }  -
 ( ( \Lambda_{ V_0} )_{ \ls[12] } )^2 \ge 0 \cr  
&& ( \Lambda_H )_{ \ls [ 11] }  ( ( \Lambda_H )_{ \ls [ 22 ] } ( \Lambda_H )_{ \ls [  33 ] }
- (( \Lambda_H)_{ \ls[23]} )^2  ) 
-  ( \Lambda_H )_{ \ls [ 12 ] }   (    ( \Lambda_H )_{ \ls [ 12] } ( \Lambda_H )_{ \ls [3 3] }  
    - ( \Lambda_H )_{ \ls [ 23] } ( \Lambda_H )_{ \ls [ 13 ] } )    \cr 
    && 
+ ( \Lambda_H )_{ \ls [ 13] }      ( ( \Lambda_H )_{ \ls [ 12] }( \Lambda_H )_{ \ls [ 23] } 
- ( \Lambda_H )_{ \ls [ 13] } ( \Lambda_H )_{ \ls [ 22 ] }
  )  \ge 0 
    \cr 
&& \Lambda_{ V_2 } \ge 0  \cr 
&& \Lambda_{ V_3} \ge 0 \, . 
\eea

With these linear and quadratic expectation values of representation theoretic matrix variables $S$  available, the expectation value of 
a general  polynomial function of $M_{ ij}$  can be expressed in terms of finite sums of products involving these 
linear and quadratic expectation values. This is an application of Wick's theorem in the context of QFT.  
We will explain this for the integrals at hand in  Appendix \ref{app:Wick} and describe the consequences of Wick's theorem explicitly for  expectation values of functions up to quartic in the matrix variables. 
We will be particularly interested in the expectation values of polynomial functions of the $M_{ ij}$ 
which are  invariant under $S_D$ action and can be parametrised by graphs. While the mixing between 
between the $S$ variables in the quadratic action is simple, there are  non-trivial  couplings 
between the  $D^2$ variables $M_{ij}$ if we expand the action in terms of the $M$ variables. 
This will lead to non-trivial expressions for the expectation values of the graph-basis polynomials.  

These expectation values were computed for the $5$-parameter Gaussian model in \cite{LMT}. 
They were referred to as theoretical expectation values $ \langle f(M) \rangle $, which were compared with 
experimental expectation values $ \langle f(M) \rangle_{ EXPT} $. These experimental expectation values were calculated by considering a list of words labelled by an index $A$ ranging from $ 1 $ to $N$, and their corresponding matrices 
$M^{A}$, 
\bea 
\langle f (M ) \rangle_{ EXPT  } = { 1 \over N } \sum_{ A =1}^{ N } f  ( M^A ) \, . 
\eea
 
We will now proceed to explicitly apply Wick's theorem to calculate the 
expectation values of permutation invariant functions labelled by graphs
for the case of quadratic functions (2-edge graphs), cubic (3-edge graphs) and quartic 
functions (4-edge graphs). We will leave the comparison of the results of this 13-parameter Gaussian 
 model to linguistic data for the future.

\section{ Graph basis invariants in terms of rep theory parameters }\label{sec:reptograph} 

In  the graph theoretic description of $S_D$ invariants constructed from $M_{ij}$, nodes 
in the graph correspond to indices, $M$ corresponds to directed edges.
At linear order we have the one-node invariant $ \sum_i M_{ ii}$ and the two-node invariant 
$ \sum_{ i , j } M_{ ij}$. At quadratic order in $M$ we have up to three nodes.  In this section, we calculate the expectation values of the linear and quadratic invariants in the Gaussian model defined 
in Section \ref{sec:IntroPIMT}. The quadratic expectation  values show non-trivial mixing between the different $M_{ij}$, unlike the 5-parameter model, where the $M_{ii}$ are decoupled from the off-diagonal elements and from each other. In that simple model, $M_{ij}$ only mixes with $M_{ji}$. 
Here the mixings are more non-trivial, but controlled by $S_D$ representation theory. 

The quantity $ F( i, j)$ defined in (\ref{asumCC})
\bea\label{natToHook} 
 F ( i , j ) = \sum_{ a } C_{ a , i} C_{ a  , j  } =  \left ( \delta_{ ij} - { 1 \over D } \right ) 
\eea
will play  an important role in the following. Its meaning is that it  is the projector for the hook representation in the natural representation. 
Deriving expressions for expectation values of permutation invariant polynomial functions 
of the matrix variable $M$ amounts to doing appropriate sums of products of $F$ factors, 
 with the arguments of these $F$ factors being related to each other according to the nature of 
 the polynomial under consideration.  In terms of the variables 
\bea 
\widetilde \mu_{ \alpha  } = \sum_{ b=1 }^2 ( \Lambda_{V_0}^{-1})_{ \alpha  \beta  } \mu_{ \beta }  \nnm 
\eea
defined in Section \ref{sec:IntroPIMT}, repeated here for convenience,
\bea
&& \langle  S^{  V_0 ; 1   } \rangle  = \wmuo \cr  
&& \langle S^{  V_0 ; 2  } \rangle = \wmut \, . 
\eea
Using the expansion (\ref{expMS}) of the matrix variables 
in terms of the rep-theoretic $S$ variables  and the 1-point functions of these in (\ref{V0tad}),(\ref{V0tad1})
and (\ref{H23tad}), we have  
the 1-point function for the matrix variables  
\bea\label{1ptanswerMij}  
\langle M_{ ij} \rangle =  { \wmuo \over D }  + {\wmut \over \sqrt  { D -1  } } F ( i , j ) \, . 
\eea

Using (\ref{quadWick1}) (\ref{quadWick2}) along with the expansion of $M$ in terms of $S$ variables (equation \ref{expMS}), 
and the two-point functions of the $S$-variables, we have 
\bea\label{conplusdiscon}  
\langle M_{ ij} M_{ kl} \rangle = \langle M_{ ij} M_{ kl} \rangle_{ \ls[\con] } + \langle M_{ ij} \rangle \langle M_{ kl} \rangle 
\eea
where 
\bea\label{Master1} 
&& \langle M_{ ij} M_{ kl} \rangle_{ \ls [ \con ] }  
= { 1 \over D^2 } \langle S^{  V_0 ; 1   } S^{  V_0 ; 1  } \rangle_{ \ls [ \con ] }   
  + { 1 \over D-1  } \sum_{ a_1 , a_2 =1 }^{ D-1}  C_{  a_1  , i  } C_{   a_1 ,   j } C_{  a_2 ,  k  } C_{   a_2 ,  l  }
 \langle  S^{ V_0 ; 2 }  S^{  V_0 ; 2  } \rangle_{ \ls [ \con ] }   \cr 
&&  + { 1 \over D \sqrt { D-1 }   } \sum_{ a =1}^{ D-1}   \langle S^{  V_0 ; 1  } S^{  V_0 ; 2   } \rangle_{ \ls[ \con ]  }  C_{  a ,  k  } C_{   a ,  l  }  +  { 1 \over D \sqrt { D-1 }   } \sum_{ a =1 }^{ D-1}   \langle S^{  V_0 ; 2   } S^{   V_0 ; 1   } \rangle_{ \ls [ \con ] }  C_{  a , i  } C_{   a , j  }   \cr 
&& +  { 1 \over D  } \sum_{ a_1,a_2 =1 }^{ D-1}  C_{ a_1 , j} C_{ a_2 , l } 
 \langle S^{  H ; 1  }_{a_1} S^{  H ; 1  }_{a_2} \rangle_{ \ls [  \con ]  }  
 + { 1 \over D }  \sum_{ a_1,a_2 =1 }^{ D-1}  C_{ a_1 ,  i} C_{ a_2 ,  k } 
 \langle S^{  H ; 2  }_{a_1} S^{  H ; 2  }_{a_2} \rangle_{  \ls [ \con ]  }  \cr 
 && 
 +  \sum_{ a_1, b_1, c_1 ,  a_2 , b_2 , c_2   =1 }^{ D -1 } C_{ a_1 , i } C_{ b_1  , j }   C_{ a_2 ,   k } C_{ b_2 ,   l }  
  C_{ a_1 , b_1 ; ~c_1 }^{ HH \rightarrow H }C_{ a_2 , b_2 ;~ c_2 }^{ HH \rightarrow H }
 \langle  S^{  H ; 3  }_{ c_1 } S^{  H ; 3  }_{ c_2 }  \rangle_{ \ls [ \con ]  }   \cr 
 && + { 1 \over D  }  \sum_{ a_1,a_2 =1 }^{ D-1}  C_{ a_1   , j} C_{ a_2 ,  k } 
 \langle S^{  H ; 1  }_{a_1} S^{  H ; 2  }_{a_2} \rangle_{ \ls [  \con ]  }  
 + { 1 \over D } \sum_{ a_1,a_2 =1 }^{ D-1}  C_{ a_1 ,   i} C_{ a_2  , l } 
 \langle S^{  H ; 2 }_{a_1} S^{  H ; 1  }_{a_2} \rangle_{  \ls [\con]  }  \cr 
 &&  +  { 1 \over \sqrt { D } } \sum_{ a_1   =1 }^{ D-1} \sum_{ a_2, b_2 , c_2 =1 }^{ D -1 }  C_{ a_1 ,  j}
  C_{ a_2  ,  k } C_{ b_2  ,  l  }    C_{ ~a_2 , ~b_2 ~;~  c_2 }^{ HH \rightarrow H }  
 \langle S^{  H ; 1  }_{a_1}    S^{  H ; 3  }_{ c_2 } \rangle_{ \ls [ \con ] }   \cr 
 && +  { 1 \over \sqrt { D } } \sum_{ a_1,b_1 , c_1  =1 }^{ D-1} \sum_{ a_2 =1 }^{ D -1 } 
 C_{ a_1 ,   i } C_{ b_1 ,   j  }    C_{ a_1 , b_1  ;~ c_1 }^{ HH \rightarrow H }  C_{ a_2 ,  l }  
 \langle S^{  H ; 3  }_{c_1}    S^{ H ; 1  }_{ a_2 } \rangle_{  \ls [ \con ] }   \cr 
 &&  +   { 1 \over \sqrt { D } } \sum_{ a_1 , c_1  =1 }^{ D-1} \sum_{ a_2, b_2 , c_2 =1 }^{ D -1 }   C_{ a_1 ,  i }   C_{ a_2  ,  k } C_{ b_2  ,  l   }    C_{ a_2 , b_2 ;~ c_2 }^{ HH \rightarrow H }
 \langle S^{  H ; 2  }_{a_1}    S^{  H ; 3  }_{ c_2 } \rangle_{ \ls [ \con ] }   \cr 
 && +  { 1 \over \sqrt { D } } \sum_{ a_1,b_1  , c_1 =1 }^{ D-1} \sum_{ a_2 =1 }^{ D -1 } C_{ a_1  ,  i } 
 C_{ b_1  ,  j  }    C_{ a_1 , b_1 ;~  c_1 }^{ HH \rightarrow H }  C_{ a_2 ,  k   }  
 \langle S^{  H ; 3  }_{c_1}    S^{  H ; 2  }_{ a_2 } \rangle_{ \ls [ \con ] }   \cr 
 && + \sum_{ a_1 , b_1    =1 }^{ D -1 }\sum_{ a_2 , b_2    =1 }^{ D -1 }  \sum_{ c_1 , c_2 =1}^{ \Dim V_2 }   C_{ a_1 ,  i } C_{ b_1 ,   j }    C_{ a_1  , b_1 ;~  c_1 }^{ HH \rightarrow V_2 }
  C_{ a_2  ,  k } C_{ b_2 , l  }    C_{ a_2 ,  b_2 ;~ c_2 }^{ HH \rightarrow V_2 }
\langle  S^{  V_2 }_{ c_1 }  S^{  V_2 }_{ c_2 } \rangle_{ \ls [ \con ] }    \cr 
&& +  \sum_{ a_1 , b_1   =1 }^{ D -1 }\sum_{ a_2 , b_2    =1 }^{ D -1 }  \sum_{ c_1 , c_2 =1}^{ \Dim V_3}   C_{ a_1  ,  i } C_{ b_1 ,  j }    C_{ ~a_1 , ~b_1  ~;~ c_1 }^{ HH \rightarrow V_3 }
  C_{ a_2 ,  k } C_{ b_2 ,  l  }    C_{ a_2 , b_2 ;~  c_2 }^{ HH \rightarrow V_3 }
\langle  S^{  V_3 }_{ c_1 }  S^{  V_3 }_{ c_2 } \rangle_{ \ls [\con  ] }  \, .
\eea

All the terms can be expressed in terms of the $F$-function defined in  (\ref{natToHook})
\bea\label{Master1pt1}  
&& \langle M_{ ij} M_{ kl } \rangle_{ \ls[ \con ] }  = {1 \over D^2 } ( \Lambda_{V_0}^{-1})_{ \ls[11]} 
+ { ( \Lambda_{V_0}^{-1})_{ \ls[22]} \over ( D-1) } F ( i , j ) F ( k ,l )  
+  {   ( \Lambda_{V_0}^{-1})_{ \ls[12]} \over D \sqrt { D-1 }    }  \left (   F ( k , l ) + F ( i , j )  \right )   \cr 
&&  + { ( \Lambda_H^{-1} )_{ \ls[11] }  \over D } F ( j , l ) + { ( \Lambda_H^{-1} )_{ \ls[22]}   \over D } F ( i , k ) + { D ( \Lambda_H^{-1} )_{ \ls[33]} \over ( D-2) } \sum_{ p , q =1}^D  F ( i , p ) F ( j , p ) F ( k , q ) F ( l , q ) F ( p , q ) \cr 
&& +   { ( \Lambda_H^{-1} )_{ \ls[12]}  \over D }  \left (  F ( j , k ) + F (  i, l )              \right )    + { ( \Lambda_H^{-1} )_{ \ls[13] }  \over \sqrt{ D-2} }  \left ( \sum_{ p =1}^D  F ( j,p) F ( k,p) F ( l , p )  + F ( i, p ) F ( j , p )F ( l , p ) \right )  \cr 
&&  + { ( \Lambda_H^{-1} )_{ \ls[23] }  \over \sqrt{ D-2} } \left ( \sum_{ p =1 }^D  F ( i , p ) F ( k , p ) F ( l , p ) 
+ F ( i , p ) F ( j , p ) F ( k , p )    \right )    \cr 
&&  + ( \Lambda_{ V_2}^{-1} ) \biggl ( { 1 \over 2 } F ( i , k ) F ( j , l ) + { 1 \over 2 } F ( i , l ) F ( j , k ) 
- { D \over D-2 } \sum_{ p , q =1}^D F ( i , p ) F (  j , p ) F ( k , q ) F ( l , q ) F ( p , q )  \cr 
&& \hspace*{4cm}  - { 1 \over ( D -1 ) } F ( i , j ) F ( k , l )  \biggr ) \cr 
&&   +    { ( \Lambda_{ V_3 }^{-1} ) \over 2  } \left ( F  ( i , k ) F ( j , l ) - F ( i , l ) F ( j , k )   
\right ) \, .  \eea
We will refer to the terms depending on $ \Lambda_{V_0} $ as $V_0$-channel 
contributions to the 2-point functions, those on $ \Lambda_H$ as $V_H$-channel (or $H$-channel) contributions, 
those on $\Lambda_{ V_2} $ as $V_2$-channel and those on $ \Lambda_{ V_3}$ as $V_3$-channel 
contributions. It will be convenient to denote these different channel contributions 
as $ \langle M_{ ij} M_{ kl} \rangle_{ V}  $ where $ V \in \{ V_0 , V_H , V_2 , V_3 \} $, so that we have 
\bea\label{Vdecomp}
\langle M_{ ij} M_{ kl} \rangle_{\ls [ \con ] }  = \langle M_{ ij} M_{ kl} \rangle^{ V_0}_{ \ls [ \con ] } 
 +  \langle M_{ ij} M_{ kl} \rangle^{ V_H }_{ \ls [ \con ] }   
 +  \langle M_{ ij} M_{ kl} \rangle^{ V_2}_{  \ls [ \con ] }  +  
  \langle M_{ ij} M_{ kl} \rangle^{ V_3}_{ \ls [  \con ] } \, . 
\eea

In arriving at the expressions for the last two terms in (\ref{Master1pt1}),
 we used the fact that these terms in (\ref{Master1})
can be expressed as 
\bea\label{easyProj} 
\Lambda_{ V_2}^{ -1}  ~~( e_i \otimes e_j , P^{ V_{ D  } \otimes V_{ D  } \rightarrow V_2} ( e_k \otimes e_l ) )  
+ \Lambda_{ V_3}^{ -1 }  ~~ ( e_i \otimes e_j , P^{ V_{ D  } \otimes V_{ D  } \rightarrow V_3 } ( e_k \otimes e_l ) )  \, . 
\eea
Here $ P^{ V_{ D } \otimes V_{ D  } \rightarrow V_2}$ is the projector from 
the tensor product of natural reps. to $V_2$, the irrep of $S_D$ associated with 
Young diagram $ [D-2,2]$. Similarly for $V_3 = [ D- 2,1,1]$. 
Now it is useful to observe that $V_3$ is just the anti-symmetric part of $ V_{ H} \otimes V_H$. 
The symmetric part decomposes as $ V_0 \oplus V_H \oplus V_2 $. 
The projector for  $V_0$ is 
\bea 
(P^{ H H \rightarrow V_0 } )_{ a_1 b_1 ; a_2 b_2 } = { 1 \over D -1  }  \delta_{ a_1 a_2 } \delta_{ b_1 b_2 } \, . 
\eea
For $V_H$ it is 
\bea 
(P^{ HH \rightarrow H } )_{ a_1 b_1 ; a_2 b_2 } =  \sum_{ c =1 }^{ D-1}  C^{ H , H \rightarrow H }_{ a_1 , b_1 ;~ c }  C^{ H , H \rightarrow H }_{ a_2 , b_2  ; ~ c }  = { D \over D -2 } \sum_{ c =1 }^{ D-1}  C_{ a_1 , b_1 ,  c } C_{ a_2 ,  b_2  , c } \, . 
\eea 
The factor $ {D \over D -2 } $ is explained in Appendix \ref{normClebschs}.

\subsection{ Calculation of $ \sum_{ i , j }  \langle M_{ ij} M_{ ij} \rangle $ } 

Following (\ref{Vdecomp}) the expectation value $ \langle M_{ ij} M_{ ij} \rangle $ 
can be written as a sum over $V$-channel contributions, where $V$ ranges over the four irreps.

\subsubsection{ Contributions from $V_2, V_3$ } 

From (\ref{easyProj}) we find 
\bea 
&& \sum_{ i , j } \langle M_{ ij} M_{ ij} \rangle_{ \ls [ \con ]  }^{ V_2 }  
= ( \Lambda_{V_2} )^{ -1} \tr_{ V_{ D } \otimes V_{ D } } ( P^{ V_{ D } \otimes V_{ D } \rightarrow V_2} )  \cr 
&& = ( \Dim ~    V_{2} ) 
 ( \Lambda_{V_2} )^{ -1} = {  D  ( D - 3 ) \over 2 }  ( \Lambda_{V_2} )^{ -1}  \, . 
\eea
The projector has eigenvalue $1$ on the subspace transforming in the irrep $V_2$ and zero elsewhere, hence the $( \Dim ~  V_2 ) $.  Similarly
\bea 
 \sum_{ i , j } \langle M_{ ij} M_{ ij} \rangle_{ \ls [ \con ]  }^{ V_3 }   
= ( \Dim ~  V_3 ) ( \Lambda_{V_3} )^{ -1} = { ( D-1)   ( D - 2  ) \over 2 } ( \Lambda_{V_3} )^{ -1}
\, . 
\eea 

Since $V_2, V_3 $ appear inside the $ V_H \otimes V_H$ subspace of $ V_{ \nat } \otimes V_{ \nat } $, we can also write the trace in $ V_{ H } \otimes V_{ H} $ and express this in terms of 
irreducible characters  
\bea 
&& tr_{ V_H  \otimes V_H } P_{ V_2} \cr 
 && = {  ( \Dim ~  V_2 )   \over D! } \sum_{ \sigma \in S_D } \chi_{ V_2} ( \sigma ) 
  \chi_H ( \sigma ) \chi_H ( \sigma ) \cr 
  && = ( \Dim ~  V_2 ) \, . 
\eea
In the last line we have used the fact that the Kronecker coefficient
for $ V_{ H } \otimes V_{ H} \rightarrow V_2 $ is $1$.

\subsubsection{Contribution from $V_H$ channel  } 

The $ ( \Lambda_H^{-1} )_{ \ls[11]} $ contribution  is 
\bea 
(  \Dim ~ V_H )  ( \Lambda_H^{-1} )_{ \ls [ 11 ]  } = ( D -1 ) ( \Lambda_H^{-1} )_{ \ls [ 11 ]  } 
\eea
which can be obtained easily from (\ref{Master1}) or (\ref{Master1pt1}). 
Similarly, the $ ( \Lambda_H^{-1} )_{ \ls[12]} $ contribution is zero. 
And the $ ( \Lambda_H^{-1})_{ \ls [ 22 ] } $ contribution is 
\bea 
 ( \Dim ~ V_H )  ( \Lambda_H^{-1} )_{22 } = ( D -1 ) ( \Lambda_H^{-1} )_{22 } \, . 
\eea  

From either (\ref{Master1}) or (\ref{Master1pt1}) we easily conclude 
that the $ (\Lambda_{ H})^{ -1}_{ \ls[13]}  , (\Lambda_{ H})^{ -1}_{ \ls[23]} $ contributions vanish. 
Starting from (\ref{Master1}), we make repeated use of  (\ref{isumC}). 

The $ ( \Lambda_{ H}^{ -1} )_{ \ls[33] } $ contribution is 
\bea 
( \Lambda_{ H}^{ -1} )_{ \ls[33] } {\tr}_{ V_H \otimes V_H} ( P^{ HH \rightarrow H} ) 
= ( \Lambda_{ H}^{ -1} )_{ \ls[33] }  ( \Dim ~ V_{ H } ) 
 = ( \Lambda_{ H}^{ -1} )_{ \ls[33] } ( D-1 )  \, . 
\eea
We used the second equation in  (\ref{ortho}). 

\subsubsection{ Contribution from $V_0$ channel } 

This is 
\bea 
\sum_{ i  , j } \langle M_{ ij} M_{ ij} \rangle_{\ls [ \con ] }^{ V_0 }  
&& = ( \Lambda_{V_0}^{-1} )_{ \ls [ 11 ] } + { 1 \over \Dim ~ V_H}  ( \Lambda_{V_0}^{-1} )_{ \ls [ 22 ]  }  \Dim ~ V_H  \cr 
 && =  ( \Lambda_{V_0}^{-1})_{ \ls [ 11] }  +  ( \Lambda_{V_0}^{-1} )_{ \ls [ 22 ] }  \, . 
\eea

\subsubsection{ Summing all channels } 

\bea\label{resultMijMijcon}  
 \sum_{ i , j } \langle M_{ ij} M_{ ij} \rangle_{  \con  }  
& = &    ( \Lambda_{V_0}^{-1})_{ \ls[11] }  +  ( \Lambda_{V_0}^{-1} )_{ \ls[22] }  
 + ( D -1 ) ( \Lambda_H^{-1} )_{ \ls[22]  } + ( D-1 )  ( \Lambda_{ H}^{ -1} )_{ \ls[33] } 
+  ( D -1 ) ( \Lambda_H^{-1} )_{ \ls[11]  }  \cr 
&& +   {  D  ( D - 3 ) \over 2 }  ( \Lambda_{V_2} )^{ -1} + { ( D-1)   ( D - 2  ) \over 2 } ( \Lambda_{V_3} )^{ -1} \, . 
\eea 
 
 The disconnected piece is 
 \bea\label{MijMijdisc}  
 && \sum_{ i , j } \langle M_{ ij} \rangle ~  \langle  M_{ ij } \rangle 
  = \sum_{ i , j }  \left ( { \wmuo \over D }  + {\wmut \over \sqrt  { D -1 } } F ( i , j )  \right )^2 \cr 
  && = \wmuo^2 +  \wmut^2    \, . 
 \eea
 and  using (\ref{conplusdiscon}) 
 \bea
&&  \sum_{ i , j } \langle M_{ ij} M_{ ij} \rangle = \wmuo^2 +  \wmut^2  
 +  \sum_{ i , j } \langle M_{ ij} M_{ ij} \rangle_{  \con  }  \cr 
 &&   = \wmuo^2 +  \wmut^2   + ( \Lambda_{V_0}^{-1})_{ \ls[11] }  +  ( \Lambda_{V_0}^{-1} )_{ \ls[22] }  
 + ( D -1 ) ( \Lambda_H^{-1} )_{ \ls[22]  } + ( D-1 )  ( \Lambda_{ H}^{ -1} )_{ \ls[33] } 
+  ( D -1 ) ( \Lambda_H^{-1} )_{ \ls[11]  }  \cr 
&& +   {  D  ( D - 3 ) \over 2 }  ( \Lambda_{V_2} )^{ -1} + { ( D-1)   ( D - 2  ) \over 2 } ( \Lambda_{V_3} )^{ -1} \, . 
 \eea

\subsection{ Calculation of $ \sum_{ i , j }  \langle M_{ ij} M_{ ji } \rangle $ } 

As in (\ref{Vdecomp}) 
 the expectation value $ \langle M_{ ij} M_{ ij} \rangle $ 
can be written as a sum over $V$-channel contributions, where $V$ ranges over the four irreps.

\subsubsection{ Contribution from multiplicity $1$ channels $V_2, V_3$ } 

  The $  ( \Lambda_{V_2}^{-1} )$ contribution is 
\bea 
&& ( \Lambda_{V_2}^{-1} ) {\tr}_{ V_H  \otimes V_H } (  P^{ V_2} \tau ) \cr  
&& = ( \Lambda_{V_2}^{-1} ) d_{ V_2 } = ( \Lambda_{V_2}^{-1} ) {  D ( D-3 ) \over 2 }  \, . 
\eea
$ \tau $ is the swop which acts on the two factors of $V_H$. We have used the fact that $ V_2$ appears in the symmetric part of $ V_{ H } \otimes V_H$. 

The $ ( \Lambda_{V_3}^{-1} )$ contribution is 
\bea 
- ( \Lambda_{V_3}^{-1} ) { ( D -1 ) ( D-2 ) \over 2 }  \, . 
\eea 
We use the fact that $ V_3$ is the antisymmetric part of $ V_H \otimes V_H$. 

\subsubsection{ Contribution from $V_H$ channel } 

The $ ( \Lambda_{ H}^{ -1 } )_{ \ls[11]} $ contribution 
From (\ref{Master1}), we have 
\bea 
{ 1 \over D } \sum_{ a_1 , a_2  } \sum_{ i , j }  C_{ a_1 ,  j } C_{ a_2  , i }  ( \Lambda_{ H}^{ -1 } )_{ \ls[11]} \delta_{ a_1 a_2} = { 1 \over D } ( \Lambda_{ H}^{ -1 } )_{ \ls[11]} \sum_{ a } 
C_{ a ,  i } C_{ a ,  j  } 
=0 
\eea
using (\ref{isumC}). Similarly the $ (\Lambda_{ H}^{ -1} )_{ \ls[22]} $ contribution is zero. 

The $ (\Lambda_H^{-1})_{\ls[12]} $ contribution is 
\bea 
{ 2  \over D } \sum_{ a , i ,  j  } C_{ a ,  j } C_{ a, j } (  \Lambda_H^{-1} )_{\ls[12]} 
= 2 { ( D-1)  } (  \Lambda_H^{-1} )_{\ls[12]} \, . 
\eea

The $ (\Lambda_H^{-1})_{\ls[33]} $ contribution is 
\bea 
&&  \sum_{ i , j } \sum_{ a , b , c , d, e }
  ( \Lambda_H^{-1} )_{\ls[33]}  C_{ a , i } C_{ b , j } C_{ c , j } C_{ d , i } 
C_{ ~a , ~b ~;~  e }^{ HH \rightarrow H} C_{ ~c , ~d  ~;~  e }^{ HH \rightarrow H} \cr 
&& = ( \Lambda_H^{-1} )_{\ls[33] } \sum_{ a , b , c, d , e } 
C_{ ~a , ~b ~;~  e }^{ HH \rightarrow H}  C_{ ~b , ~ a ~;~ e }^{ HH \rightarrow H}  \cr 
&& = ( \Lambda_H^{-1} )_{\ls [ 33] } {\tr}_{ \ls[H \otimes H]} P^{ H \otimes H \rightarrow H } s  = 
 ( \Lambda_H^{-1} )_{\ls [ 33] } {\tr}_{ \ls[H \otimes H]} P^{ H \otimes H \rightarrow H } ( { 1 + s \over 2 } - { 1 - \tau \over 2 } )  \cr 
&& = ( \Lambda_H^{-1} )_{\ls[33]} d_H = ( D-1)  ( \Lambda_H^{-1} )_{\ls[33]} \, . 
\eea

In the penultimate line, we have introduced the swop $ s $ which 
exchanges the two factors in $ H \otimes H$. We know that $H$ appears in the 
symmetric part of $H \otimes H$, so the swop leaves it invariant.

Use of the equation  (\ref{isumC}) shows that the $ ( \Lambda_{ H}^{ -1}  )_{ \ls[13]} , ( \Lambda_{ H}^{ -1}  )_{ \ls[23] } $ dependent terms vanish. 

Collecting all the $V_H$-channel contributions, we have 
\bea 
\sum_{ i , j } \langle M_{ ij} M_{ ji} \rangle^{ V_H}_{\ls[\con]}  
 = 2 { ( D-1) \over D } (  \Lambda_H^{-1} )_{\ls[12]} + ( D-1)  ( \Lambda_H^{-1} )_{\ls[33]} \, . 
\eea

\subsubsection{ Contribution from $V_0$ channel } 

The first term from (\ref{Master1}) is 
\bea 
(\Lambda_{ V_0}^{ -1} )_{ \ls[11] }  \, . 
\eea
The second term is 
\bea 
{ ( \Lambda_{V_0}^{ -1}  )_{ \ls[22] }  \over d_{ H } } \sum_{ i , j }  \sum_{ a_1 , a_2 } C_{ a_1 ,  i } C_{ a_1  , j } C_{ a_2 ,  j } C_{ a_2 ,  i }  = { ( \Lambda_{V_0}^{ -1}  )_{ \ls[22] }  \over D-1  } \sum_{ a_1 , a_2 } \delta_{ a_1 a_2 } \delta_{ a_1 a_2 } =   ( \Lambda_{V_0}^{ -1}  )_{ \ls[22] }   \, . 
\eea
The third term is
\bea 
{ 1 \over D \sqrt{ D-1  } } \sum_{ i , j } \sum_a C_{ a , j} C_{ a , i} ( \Lambda_{V_0}^{ -1} )_{ \ls[12] } 
\, . 
\eea
which vanishes using (\ref{isumC}). The last term vanishes for the same reason. 

So collecting the $V_0$-channel contributions to $ \sum_{ i , j } \langle M_{ ij} M_{ ji} \rangle_{ \ls [ \con ] }  $, we have 
\bea
 \sum_{ i , j } \langle M_{ ij} M_{ ji} \rangle_{ \ls [ \con ]  }   =  (\Lambda_{ V_0}^{ -1} )_{ \ls[11] }  
 + ( \Lambda_{V_0}^{ -1}  )_{ \ls[22] }   \, . 
 \eea

\subsubsection{ Summing all channels } 
\bea\label{MijMjicans}  
 && \sum_{ i , j } \langle M_{ ij} M_{ ji} \rangle_{ \ls [ \con ] }   =    ( \Lambda_{V_2}^{-1} ) { ( D  ) ( D-3 ) \over 2 } - ( \Lambda_{V_3}^{-1} ) { ( D -1 ) ( D-2 ) \over 2 }\cr 
&& 
+  ~~ 2 { ( D-1)  } (  \Lambda_H^{-1} )_{\ls[12]}  + ( D-1)  ( \Lambda_H^{-1} )_{\ls[33]}   + ~~ (\Lambda_{ V_0}^{ -1} )_{ \ls[11] }  
 + ( \Lambda_{V_0}^{ -1}  )_{ \ls[22] }   \, . 
\eea

Since $ F ( i , j ) = F ( j , i ) $, 
the disconnected piece is the same in (\ref{MijMijdisc})
\bea\label{MijMjidisc}  
&& \sum_{ i , j } \langle M_{ ij} \rangle \langle  M_{ ji} \rangle =
 \sum_{ i , j } \langle M_{ ij} \rangle \langle  M_{ ij} \rangle 
\cr 
&& =  \wmuo^2 +  \wmut^2   \, . 
\eea
and  $ \sum_{ i , j }  \langle M_{ ij} M_{ ij} \rangle $  in terms of the $ \mu , \Lambda $ parameters of 
the Gaussian model is the sum of the expressions in   (\ref{MijMjicans})
and (\ref{MijMjidisc}).

\subsection{ Calculation of $ \sum_{ i , j }  \langle M_{ ii} M_{ij } \rangle $ }\label{seciiij}  

An important observation here is that the sum over $j$ projects the representation $V_D$  to the trivial 
irrep  $V_0$, which follows from the formula for $ C_{0i}$ in (\ref{theCs}).
This means that when we expand $M_{ii} $ and $M_{ ij} $ into $S$ variables 
as in the first line of  (\ref{MtoS}), we only need to keep the term $S^{H0}$ 
or $S^{ 00} $ from the expansion of $M_{ ij}$. 

\subsubsection{ Contribution from $V_2, V_3$ channels } 

From the above  observation, and since $V_2 , V_3$ appear only in $\langle S^{HH} S^{ HH} \rangle  $, 
we immediately see that 
\bea 
\sum_{ i , j } \langle M_{ ii} M_{ ij} \rangle^{ V_2 }_{ \ls [ \con ]  }  = 
\sum_{ i , j } \langle M_{ ii} M_{ ij} \rangle^{ V_3 }_{ \ls [ \con ]  }   = 0 \, . 
\eea

\subsubsection{ Contribution from $V_H$ channel } 

From the above observation, the only non-zero contributions in the $V_H$ channel 
come from $ \langle S^{ HH \rightarrow H} S^{ H0 } \rangle $, $ \langle S^{ H0} S^{ H0} \rangle $ 
and $ \langle S^{ 0H} S^{ H0} \rangle $. These are 
\bea 
{ 1 \over D } \sum_{ a_1 , a_2 , i , j  } C_{ a_1 , i } C_{ a_2 , i } ( \Lambda_H^{ -1} )_{\ls[12] } 
\delta_{ a_1 a_2 }  = { ( D-1)  } ( \Lambda_H^{ -1} )_{\ls[12] } 
\eea 
\bea 
{ 1 \over D } \sum_{ a_1 , a_2 , i  , j } C_{ a_1 ,  i } C_{ a_2  , i }  ( \Lambda_H^{ -1} )_{\ls[22] }  \delta_{ a_1 , a_2 } = { ( D-1)  } ( \Lambda_H^{ -1} )_{\ls[22] }
\eea
\bea 
&& { 1 \over \sqrt { D } } \sum_{ a_1 , b_1 , a_2 , i , j  } C_{ a_1 ,  i } C_{ b_1 ,  i }
 C^{ H , H \rightarrow H}_{ a_1 , b_1  ;~  c_1 } 
 C_{ a_2  , i } \delta_{ c_1 a_2 } ( \Lambda_{ H }^{ -1} )_{ \ls[32]} \cr 
 && = { D \over \sqrt { D -2 } } \sum_{ a_1 , b_1 , a_2 , i } C_{ a_1 ,  i } C_{ b_1  , i }
 C_{ a_1 ,  b_1  , c_1 } 
 C_{ a_2  , i } \delta_{ c_1 a_2 } ( \Lambda_{ H }^{ -1} )_{ \ls[32]}  \cr
&& = { D \over \sqrt{ D -2 }  }  ( \Lambda_{ H }^{ -1} )_{ \ls[32]}  \sum_{ a_1 , b_1 , c_1 } 
                C_{ a_1 , b_1 , c_1 } C_{ a_1 , b_1 , c_1 } \cr 
 && =   { ( D-1 )\sqrt{ (D-2)   }} ~  ( \Lambda_{ H}^{-1} )_{ \ls[23]} \, . 
 \eea
 Note that $ \Lambda_H$ is a symmetric $ 3 \times 3 $ matrix 
 and  $ \Lambda_{ H}^{-1} )_{ \ls[23]} =( \Lambda_{ H}^{-1} )_{ \ls[32] }$.  In the penultimate step, we have used the normalization equation (\ref{normCabc}). 
 These add up to 
\bea 
&& \sum_{ i , j } \langle M_{ ii} M_{ ij} \rangle^{ V_H}_{ \ls[\con]}  
=  { ( D-1) } ( \Lambda_H^{ -1} )_{\ls[12] } + { ( D-1)  } ( \Lambda_H^{ -1} )_{\ls[22] }
+  ( D-1 )\sqrt{ (D-2)   } ( \Lambda_{ H}^{-1} )_{ \ls[23]} \, .  \cr 
&& 
\eea

\subsubsection{ Contribution from $V_0$ channel } 

The non-zero contributions come from $ \langle S^{ 00} S^{ 00} \rangle $ 
and $\langle S^{ HH\rightarrow 0 } S^{ 00} \rangle $. They are 
\bea 
\sum_{ i , j } \langle M_{ii} M_{ ij} \rangle_{ \ls [ \con ]  }^{ V_0 } 
&& = { 1 \over D^2 } \sum_{ i , j } ( \Lambda_{ V_0 }^{ -1 } )_{ \ls [ 11  ] } 
  + { 1 \over D  \sqrt { d_H } } \sum_{ i , j } \sum_{ a }  
  ( \Lambda_{V_0}^{ -1} )_{ \ls[12]} C_{ a , i } C_{ a , i } \cr 
  && = (\Lambda_{V_0}^{ -1})_{ \ls[11] }  + { 1 \over \sqrt { D -1  } } \sum_{ i , a } ( \Lambda_{ V_0 }^{-1} )C_{ a , i } C_{ a , i } \cr 
  && =  ( \Lambda_{V_0}^{ -1} )_{ \ls[11] }  + \sqrt { d_H } ( \Lambda_{V_0}^{-1} )_{ \ls[12]} =
  (  \Lambda_{V_0}^{ -1}  )_{ \ls[11] } + \sqrt { ( D -1 )  } ( \Lambda_{V_0}^{-1} )_{ \ls[12]}  \, . 
\eea

\subsubsection{ Summing all channels } 

\bea\label{iiijans}
\sum_{ i , j } \langle M_{ii} M_{ ij} \rangle_{ \con } 
=  
   ( \Lambda_{V_0}^{ -1} )_{ \ls[11]}  + \sqrt { ( D -1 )  } ( \Lambda_{V_0}^{-1} )_{ \ls[12]}  \cr
    + { ( D-1) } ( \Lambda_H^{ -1} )_{\ls[12] } + { ( D-1) } ( \Lambda_H^{ -1} )_{\ls[22] }
+  ( D-1 )\sqrt{ (D-2)   } ( \Lambda_{ H}^{-1} )_{ \ls[23]} \, .  \cr 
&& 
\eea

The disconnected piece is 
\bea 
 &&  \sum_{ i , j  } \langle M_{ii } \rangle \langle M_{ i j  } \rangle 
  =  \sum_{ i , j }  \left (  { \wmuo \over D }    + {  \wmut \over { \sqrt { D-1 } }     } F ( i , i )        \right ) 
    \left (     { \wmuo \over D }    + {  \wmut \over { \sqrt { D-1 } }    }   F ( i , j  )            \right )  \cr 
    &&  =   \wmuo^2   + { \wmuo \wmut \over D \sqrt { D-1 } } \sum_{ i , j } ( 1 - { 1 \over D } ) \cr 
    && =  \wmuo^2  +  { \wmuo \wmut    \sqrt { ( D-1 ) }   }
\eea 
 so we have 
\bea 
\sum_{ i , j } \langle M_{ii} M_{ ij} \rangle = \sum_{ i , j } \langle M_{ii} M_{ ij} \rangle_{ \con }
+ \wmuo^2  +  { \wmuo \wmut   \sqrt { D-1 } } \, . 
\eea
with the first term given by (\ref{iiijans})
 
\subsection{ Calculation of $ \sum_{ i , j } \langle M_{ ii} M_{ ji } \rangle  $ } 

We can write down the answer from inspection of (\ref{iiijans})
\bea 
&& \sum_{ i , j } \langle M_{ii} M_{ ji} \rangle_{ \ls [ \con ] } 
= ( \Lambda_{V_0}^{ -1} )_{ \ls[11]}  + \sqrt { ( D -1 )  } ( \Lambda_{V_0}^{-1} )_{ \ls[12]}  \cr
&&     + { ( D-1) } ( \Lambda_H^{ -1} )_{\ls[12] } + { ( D-1)  } ( \Lambda_H^{ -1} )_{\ls[11] }
+  ( D-1 )\sqrt{ (D-2)   }  ( \Lambda_{ H}^{-1} )_{ \ls[13]} \, . 
\eea

The reasoning is as follows. The sum over $j$ projects to $V_0$. 
This means that the only non-zero contributions are, from the $V_0$ channel, 
\bea 
&& \langle S^{ 00} S^{ 00} \rangle_{ \ls [ \con ]  }  \cr 
&& \langle S^{HH\rightarrow 0 } S^{ 00} \rangle_{ \ls [  \con ] }  
\eea
and from the $ V_H$ channel 
\bea 
&& \langle S^{ 0H} S^{ 0H} \rangle_{  \ls [ \con ]  }  \cr 
&& \langle S^{ H0} S^{ 0 H } \rangle_{ \ls [ \con ]  }  \cr 
&& \langle S^{ HH \rightarrow H} S^{ 0H} \rangle_{  \ls [ \con ]  }  \, . 
\eea
This identifies the contributing  entries of $ \Lambda_{ V_0}^{ -1} ,\Lambda_{ H}^{-1} $ using 
the indexing in  (\ref{convV0}) and (\ref{convVH}). given the similarity between the expectation value in 
section \ref{seciiij}, we have contributions of the same form, up to taking care of the right indices 
on $\Lambda^{ -1}_{ V_0} , \Lambda^{-1}_{ H} $.

Given the symmetry of $ F ( i , j ) $ under exchange of $ i , j $, the disconnected piece is the same as above
\bea 
&& \sum_{ i , j } \langle M_{ii} M_{ j i } \rangle
= ( \Lambda_{V_0}^{ -1} )_{ \ls[11]}  + \sqrt { ( D -1 )  } ( \Lambda_{V_0}^{-1} )_{ \ls[12]}  \cr
&&     + { ( D-1)  } ( \Lambda_H^{ -1} )_{\ls[12] } + { ( D-1)  } ( \Lambda_H^{ -1} )_{\ls[11] }
+  ( D-1 )\sqrt{ (D-2)   }  ( \Lambda_{ H}^{-1} )_{ \ls[13]}  \cr 
&& + \wmuo^2  +  { \wmuo \wmut   \sqrt { ( D-1 ) }  } \, . 
\eea

\subsection{ Calculation of $ \sum_{ i , j , k } \langle M_{ ij} M_{ik} \rangle  $ } 

The sums over $ j , k $ project to $V_0$. The non-vanishing 
contributions are 
$ \langle S^{00} S^{00} \rangle_{ \con }  $ and $  \langle S^{ H0} S^{H0} \rangle_{ \con }  $. They add up to 
\bea\label{ansijik}  
\sum_{ i , j , k } \langle M_{ ij} M_{ik} \rangle_{ \ls [  \con ]  }  
= D ( \Lambda_{V_0}^{ -1})_{ \ls[11]}   + D ( D-1) ( \Lambda_{ H}^{-1} )_{ \ls[22] } \, . 
\eea 
The disconnected part is  
\bea\label{discijik}  
\sum_{ i , j , k  } \langle M_{ ij } \rangle ~ \langle M_{ik} \rangle = { \wmuo^2  D } 
\eea
leading to 
\bea 
\sum_{ i , j , k } \langle M_{ ij} M_{ik} \rangle = 
D ( \Lambda_{V_0}^{ -1})_{ \ls[11]}   + D ( D-1) ( \Lambda_{ H}^{-1} )_{ \ls[22] }  
+ { \wmuo^2  D }  \, . 
\eea

\subsection{Calculation of $ \sum_{ i,j,k} \langle  M_{ ij} M_{ kj} \rangle  $ } 

Now we are projecting to $V_0$ on the first index of both $M$'s. 
This means that the contributing terms are 
$ \langle  S^{ 0 0 } S^{ 00 } \rangle $ 
and $ \langle S^{ 0H} S^{ 0H} \rangle $. 

Repeat the same steps as above in (\ref{ansijik}) to get 
\bea\label{ansijkj}  
 \sum_{ i,j,k} \langle M_{ ij} M_{ kj} \rangle_{ \ls[con] }   =  
  D ( \Lambda_{V_0}^{ -1})_{ \ls[11]}   + D  ( D-1)  ( \Lambda_{ H}^{-1} )_{ \ls[11] } \, . 
\eea
The only difference is that we are picking up the $(1,1)$ 
matrix  element of $( \Lambda_{ H}^{-1})$ instead of the $(2,2)$ element, since  we defined 
$ S^{ 0H} = S^{ \{ V_0 ; 1 \} } $ and  $ S^{ H0} = S^{ \{ V_0 ; 2 \} }$. 
 
Adding  the disconnected piece, which is the same as (\ref{discijik}), 
we have 
\bea 
\sum_{ i,j,k} \langle M_{ ij} M_{ kj} \rangle 
=   D ( \Lambda_{V_0}^{ -1})_{ \ls[11]}   + D  ( D-1)  ( \Lambda_{ H}^{-1} )_{ \ls[11] } 
+ \wmuo^2 D \, . 
\eea

\subsection{ Calculation of  $ \sum_{ i,j,k} \langle M_{ ij} M_{ jk} \rangle $ } 

We are now projecting to $ V_0 $ on first index of one of the matrices and second index of the other. 
Hence the contributing terms are $ \langle S^{ 00} S^{ 00} \rangle $ and 
$ \langle S^{ 0H} S^{ H0 } \rangle $. 
The result is 
\bea\label{resMijMjk}  
 \sum_{ i,j,k} \langle M_{ ij} M_{ j k } \rangle_{ \ls[\con ] }   = 
  D ( \Lambda_{V_0}^{ -1})_{ \ls[11]}   + D ( D-1)   ( \Lambda_{ H}^{-1} )_{ \ls[12] } \, . 
\eea
and 
\bea 
\sum_{ i,j,k} \langle M_{ ij} M_{ j k } \rangle =
  D ( \Lambda_{V_0}^{ -1})_{ \ls[11]}   + D ( D-1)   ( \Lambda_{ H}^{-1} )_{ \ls[12] } 
  +  \wmuo^2 D \, . 
\eea

\subsection{ Calculation of $ \sum_{i,j,k,l  } \langle  M_{ ij} M_{kl}   \rangle  $ } 

Here we project to $V_0$ on all four indices, so 
\bea 
\sum_{ i,j,k,l} \langle  M_{ ij} M_{ kl} \rangle_{ \ls[ \con ]  }  = D^2 \langle  S^{ 00} S^{00} \rangle 
 =  D^2 ( \Lambda^{-1}_{V_0} )_{ \ls[11] } \, . 
\eea
Adding the  disconnected piece we have 
\bea 
\sum_{ i,j,k,l} \langle  M_{ ij}  M_{ kl} \rangle 
=  D^2 ( \Lambda^{-1}_{V_0} )_{ \ls[11] } + { \wmuo^2  D^2  } \, . 
\eea

\subsection{ Calculation of $\sum_{ i } \langle  M_{ ii}^2 \rangle $  }

\subsubsection{The $V_0$ channel }

The contribution from the $V_0$ channel is given by 
\bea 
&& \sum_{ i } \langle  M_{ii}^2 \rangle^{V_0}_{ \ls [ \con ]   }  \cr 
 && = { 1 \over D^2 } \sum_i( \Lambda_{ V_0}^{ -1} )_{ \ls[11] } 
+  { 1 \over ( D-1 )  } \sum_{ i }  F ( i , i )^2   ( \Lambda_{ V_0 }^{ -1} )_{ \ls[22] } 
+ { 2 \over D \sqrt{ D-1  } }  \sum_{ i } F ( i , i )  ( \Lambda_{ V_0 }^{ -1} )_{ \ls[12] } \cr 
&& =  { 1 \over D } ( \Lambda_{ V_0 }^{ -1} )_{ \ls[ 11 ] } + { ( D-1) \over D  } ( \Lambda_{V_0}^{-1} )_{ \ls[22] }
+ 2 {  \sqrt { D-1}    \over D }  ( \Lambda_{V_0}^{ -1})_{ \ls[12]} \, . 
\eea

\subsubsection{ The $ V_H$ channel } 

It is convenient to use (\ref{Master1pt1}) to arrive at 
\bea 
&& \sum_{ i } \langle  M_{ii}^2 \rangle_{ \ls [ \con ] }^{ V_H }  = { D-1 \over D } (  \Lambda_{ H}^{ -1} )_{ \ls[11]}  
+ ( \Lambda_{ H}^{ -1} )_{ \ls[22]} { ( D-1) \over D }  + D^{-1} ( D-1) ( D-2)  
( \Lambda_{ H}^{ -1} )_{ \ls[33]} \cr 
&& + 2 { ( D-1) \over D } ( \Lambda_{ H}^{ -1} )_{ \ls[12]}  
 + 2 { ( \Lambda_H^{ -1} )_{ \ls[13] }  \over  D  } ( D -1  ) \sqrt { ( D -2 )}  
  + 2  { ( \Lambda_H^{ -1} )_{ \ls[23] }  \over  D   } ( D -1  ) \sqrt { ( D -2 )}  \, . \cr 
  && 
\eea
Useful equations in arriving at the above are the sums 
\bea 
&& \sum_{ i , p , q } F ( i , p ) F ( i , p ) F ( i , q ) F ( i , q ) F ( p , q ) 
= { ( D -2)^2 ( D -1 ) \over D^2 }  \cr 
&& \sum_{ i , p } ( F (  i , p ) )^3 = { (D -1 ) ( D -2 ) \over D }  \,  .
\eea
which can be obtained by hand or with the help of  Mathematica. In the latter case, it is 
occasionally easier to evaluate for a range of integer $D$ and fit to a form
 $ { Polynomial   (D )  / D^{ some ~  power } } $.

\subsubsection{The $ V_2 , V_3$ channels  }

Now calculate the $HH \rightarrow V_2$ and $HH \rightarrow V_3$ channel. 
\bea
&& \sum_{ i  } \langle  M_{ii}^2  \rangle^{V_2  }_{ \ls[ \con ] } 
= \sum_{ i  } \sum_{ a , b , c , d } 
C_{ a , i } C_{ b , i } C_{c , i} C_{d , i} C^{HH;V_2}_{ab; e } C^{HH;V_2}_{cd;f} \langle  S^{HH\rightarrow V_2 }_{e} S^{HH\rightarrow V_2}_f \rangle_{ \con }   \cr
&& = (\Lambda_{V_2}^{-1} )\sum_{ i , j } \sum_{ a , b , c , d , e  } 
C_{ a , i } C_{ b , i } \delta_{ i j } C_{c , j} C_{d  , j } C^{HH;V_2}_{~a , ~b ~;~ e }
 C^{HH;V_2}_{~c , ~d ~;~ e}  \cr
&& =  (\Lambda_{V_2}^{-1} ) \sum_{ a , b , c , d , e  } 
C_{ a , i } C_{ b , i } \left ( C_{ e  , i } C_{ e , j } + { 1 \over D}   \right )  C_{c ,  j} C_{d ,  j } C^{HH;V_2}_{~ a , ~ b ~ ; ~  e } C^{HH;V_2}_{~ c , ~d ~ ;~ e}   \cr 
&& =  { D -2 \over D } (\Lambda_{V_2}^{-1} )  tr_{ H \otimes H} ( P_H P_{V_2}  ) + { 1 \over D  }(\Lambda_{V_2}^{-1} ) \sum_{ a, b } C_{ ~a , ~a ~;~  e }^{ HH \rightarrow V_2}  C^{ HH \rightarrow V_2}_{ ~b , ~ b ~ ;~  e} \cr 
&& = { D -2 \over D } (\Lambda_{V_2}^{-1} )  tr_{ H \otimes H} ( P^{H, H \rightarrow H} P^{H , H \rightarrow V_2}  ) + { 1 \over D  }  (\Lambda_{V_2}^{-1} ) \sum_{ a, b } C_{ ~a , ~ a ~;~  e }^{ HH \rightarrow V_2} 
 C^{ HH \rightarrow V_2}_{ ~b , ~b ~;~  e} \cr 
&&   = 0  + { (\Lambda_{V_2}^{-1} ) \over D -2 } \sum_{ a , b , i , j } C_{ a , i } C_{ a , i } C_{ e , i } C_{ b , j } C_{ b , j } C_{ e , j } \cr 
&& = { (\Lambda_{V_2}^{-1} ) \over D -2 } \sum_{ i ,j } F ( i , j ) F ( i , i ) F ( j , j )  
= (\Lambda_{V_2}^{-1} ){ ( D-1)^2 \over D^2 ( D-2 ) } \sum_{ i , j } F ( i , j )  = 0 \, . 
\eea
 We used the fact that the projectors for $H , V_2$ are orthogonal.   

Similarly, the contribution from $V_3$ is zero. Another to arrive at the same answer is to 
recognise that $V_3$ is the antisymmetric part, so 
\bea 
P^{ H, H \rightarrow V_3}_{ ab; cd } = { 1 \over 2} \left ( \delta_{ ac} \delta_{ bd } - \delta_{ ad} \delta_{ bc} \right ) \, . 
\eea

\subsubsection{ Summing the channels } 
\bea\label{Miisqres} 
&& \sum_{ i } \langle M_{ ii}^2 \rangle_{ \ls [ \con ]  }  
 =   { 1 \over D } ( \Lambda_{ V_0 }^{ -1} )_{ \ls[ 11 ] } + { ( D-1) \over D  } ( \Lambda_{V_0}^{-1} )_{ \ls[22] }
+ 2 {  \sqrt { D-1}    \over D }  ( \Lambda_{V_0}^{ -1})_{ \ls[12]} \cr 
 && + { D-1 \over D } (  \Lambda_{ H}^{ -1} )_{ \ls[11]}  
+ ( \Lambda_{ H}^{ -1} )_{ \ls[22]} { ( D-1) \over D }  + D^{-1} ( D-1) ( D-2)  
( \Lambda_{ H}^{ -1} )_{ \ls[33]} \cr 
&& + 2 { ( D-1) \over D } ( \Lambda_{ H}^{ -1} )_{ \ls[12]}  
 + 2 { ( \Lambda_H^{ -1} )_{ \ls[13] }  \over  D  } ( D -1  ) \sqrt { ( D -2 )}  
  + 2  { ( \Lambda_H^{ -1} )_{ \ls[23] }  \over  D   } ( D -1  ) \sqrt { ( D -2 )} \, . \cr 
  && 
\eea

The disconnected part is 
\bea 
&& \sum_{ i } \langle M_{ ii} \rangle ~ \langle M_{ ii} \rangle 
= \sum_{ i }   \left (    { \wmuo \over D }  +  { \wmut \over \sqrt { D - 1 } }  F ( i , i ) \right )^2    \cr 
&&  = {  \wmuo^2 \over D }  + 2 \wmuo \wmut   { \sqrt { D -1 } \over D } 
+ { \wmut^2 }  { ( D-1) \over D  } \, . 
\eea

\subsection{ Calculation of  $  \sum_{ i, j }  \langle M_{ii} M_{jj} \rangle  $ }  

Since $\sum_i M_{ii} $ and $ \sum_j M_{ jj} $ are $S_D$ invariant, 
we only have contributions from the $V_0$ channel. Use the first four terms of 
(\ref{Master1}) to get 
\bea 
\sum_{ i , j } \langle M_{ ii} M_{ jj} \rangle_{ \ls [ \con ] }  
=  ( \Lambda_{V_0}^{-1} )_{11} + ( D-1 )  ( \Lambda_{V_0}^{-1} )_{22} + 2  \sqrt{ D-1  }    ( \Lambda_0^{-1} )_{12} \, . 
\eea 
Using 
\bea 
\sum_{ i }   \left ( { \wmuo \over D } + { \wmut \over  \sqrt { D -1 } } F ( i , i )  \right ) 
= \wmuo + \wmut \sqrt { D -1 } 
\eea
the disconnected part is 
\bea 
\sum_{  i , j  } \langle M_{ ii} \rangle \langle M_{ jj} \rangle 
= \wmuo^2  + 2 \wmuo \wmut { \sqrt { D -1 }  }  + \wmut^2 ( D -1 ) 
\eea
so that 
\bea 
\sum_{ i , j } \langle M_{ ii} M_{ jj} \rangle
= \sum_{ i , j } \langle M_{ ii} M_{ jj} \rangle_{ \ls [ \con ] }  + \wmuo^2  + 2 \wmuo \wmut { \sqrt { D -1 }  }  + \wmut^2 ( D -1 ) \, . 
\eea

\subsection{ Calculation of  $ \sum_{ i , j , k } \langle M_{ ii} M_{ jk} \rangle  $ } 

Here we get contributions from $ \langle S^{ HH \rightarrow 0 } S^{ 00} \rangle_{ \ls [ \con ] }  $
 and $\langle S^{ 00} S^{ 00} \rangle_{ \ls [ \con ] }  $. Adding these up from (\ref{Master1})
\bea 
\sum_{ i , j , k } \langle M_{ ii} M_{ jk} \rangle_{ \ls [ \con ] }   
= D ( \Lambda_{ V_0}^{-1} )_{ \ls[11] } + { D \sqrt{  D -1 } }  ( \Lambda_{V_0}^{-1} )_{\ls [ 12 ] } 
\, . 
\eea

The disconnected part is 
\bea 
\sum_{ i , j , k } \langle M_{ ii} \rangle ~ \langle M_{ jk } \rangle 
=  ( \wmuo + \wmut \sqrt { D -1 }  )   ~  D \wmuo 
\eea
hence 
\bea 
\sum_{ i , j , k } \langle M_{ ii} M_{ jk} \rangle
= D ( \Lambda_{ V_0}^{-1} )_{ \ls[11] } + D \sqrt{  D -1 }   ( \Lambda_{V_0}^{-1} )_{\ls [ 12 ] }  
 +  ( \wmuo + \wmut \sqrt { D -1 }  )   ~  D \wmuo \, . 
\eea

\subsection{ Summary of results for quadratic expectation values   in a  large $D $ limit }\label{sec:largeD}

It is interesting to collect the results for the connected quadratic expectation values and consider the large $D$ limit. Let us assume that all the $ \Lambda_{ V_0} , \Lambda_{ H} , \Lambda_{ V_2 } , \Lambda_{ V_3}  $ scale in the same way as $ D \rightarrow \infty $ and consider the sums normalized by the appropriate factor of $D$  
\begin{align*} 
& { 1 \over D^2 } \sum_{ i , j } \langle M_{ ij} M_{ ij} \rangle_{ \con } 
  =            {  1 \over 2 }  \left ( ( \Lambda_{V_2} )^{ -1} +  ( \Lambda_{V_3} )^{ -1} \right ) \cr
& { 1 \over D^2 } \sum_{ i , j } \langle M_{ ij} M_{ ji } \rangle_{ \con } 
 =  {  1 \over 2 }  \left ( ( \Lambda_{V_2} )^{ -1} -   ( \Lambda_{V_3} )^{ -1} \right )  \cr 
& { 1 \over D^2 }  \sum_{ i , j } \langle M_{ii} M_{ ij} \rangle_{\con} 
=   { 1 \over  D^{3/2}     } ( \Lambda_{V_0}^{-1} )_{ \ls[12]}  
+ { 1 \over D^{3/2} } ( \Lambda_{ H}^{-1} )_{ \ls[23]}  \cr 
& { 1 \over D^2} \sum_{ i , j } \langle M_{ii} M_{ ji} \rangle_{\con} 
=   { 1 \over D^{ 3/2 } }  ( \Lambda_{V_0}^{-1} )_{ \ls[12]}  
+ { 1 \over D^{ 3/2}  }  ( \Lambda_{ H}^{-1} )_{ \ls[13]} \cr 
& { 1 \over D^3} \sum_{ i , j , k } \langle M_{ ij} M_{ik} \rangle_{ \con} 
= { 1 \over D }  ( \Lambda_{ H}^{-1} )_{ \ls[22] }  \cr 
& { 1 \over D^3} \sum_{ i,j,k} \langle M_{ ij} M_{ kj} \rangle_{ \con }   =  
    { 1 \over D} ( \Lambda_{ H}^{-1} )_{ \ls[11] } \cr 
&{ 1 \over D^3}  \sum_{ i,j,k} \langle M_{ ij} M_{ j k } \rangle_{ \con}   = 
   { 1 \over D }   ( \Lambda_{ H}^{-1} )_{ \ls[12] } \cr 
   & { 1 \over D^4} \sum_{ i,j,k,l} \langle  M_{ ij} M_{ kl} \rangle_{ \con}  
 =  { 1 \over D^2}  ( \Lambda^{-1}_{V_0} )_{ \ls[11] }  \cr 
 & { 1 \over D } \sum_{ i } \langle M_{ ii}^2 \rangle_{ \con}  
  = ( \Lambda_{ H}^{ -1} )_{ \ls[33]}  \cr 
 & { 1 \over D^2} \sum_{ i , j }  \langle M_{ ii} M_{ jj} \rangle_{ \con} 
=  { 1 \over D }  ( \Lambda_{V_0}^{-1} )_{\ls[11] }   +  { 1 \over D }  ( \Lambda_{V_0}^{-1} )_{\ls[22] } \cr 
& { 1 \over D^3}  \sum_{ i , j , k } \langle M_{ ii} M_{ jk} \rangle_{ \ls[\con] }   
= { 1 \over D^2}  ( \Lambda_{ V_0}^{-1} )_{ \ls[11] }   \, . 
  \end{align*} 
  The dominant expectation values in this limit are the first, second and ninth. 
  These are the quadratic expressions which enter the simplified 5-parameter model considered in \cite{LMT} (see Equation (\ref{themodel})). It will be interesting to systematically  explore the different large $D$ scalings of the parameters in real world data, e.g. the computational linguistics setting of \cite{LMT} or in any other situation where permutation invariant matrix Gaussian matrix distributions can be argued to be appropriate.

\section{ A selection of cubic expectation values }\label{sec:cubic}  

In this section we use Wick's theorem from Appendix \ref{app:Wick} to express 
expectation values of cubic functions of matrix variables in terms
of linear and quadratic expectation values. The permutation invariance condition requires sums 
of indices over the range $ \{ 1 , \cdots , D \}$. This leads to non-trivial sums of products of 
the natural-to-hook projector $ F ( i , j ) $. The invariants at cubic order are $52$ in number (Appendix B of \cite{LMT})  and correspond to graphs  with up to $6$ nodes. 

\subsection{ 1-node case $ \sum_i \langle  M_{ii}^3 \rangle  $ } 

Using \ref{cubicWick}, we have 
\bea 
\sum_{ i } \langle M_{ii}^3 \rangle  = 3 \sum_{ i } \langle  M_{ ii}^2 \rangle_{\con}  ~ \langle M_{ ii} \rangle + \langle M_{ ii} \rangle^3  \, . 
\eea
Specialising  (\ref{1ptanswerMij})
\bea 
\langle M_{ ii} \rangle = { 1 \over D } {\widetilde \mu}_{ \ls[1] }  
  +   { \sqrt { ( D-1) }   \over D }  {\widetilde \mu}_{ \ls[2 ] }  \, . 
\eea
Since this is independent of $i$, we can use (\ref{Miisqres}) to get 
\bea 
&& \sum_{ i } \langle M_{ ii}^3 \rangle 
=  3 \left ( { 1 \over D }  {\widetilde \mu}_{ 1} 
+ { \sqrt { ( D-1) }   \over D } ) {\widetilde \mu}_{ 2}  \right )   \times \cr 
&&  \biggl ( { 1 \over D } ( \Lambda_{ V_0 }^{ -1} )_{ \ls[ 11 ] } + { ( D-1) \over D  } ( \Lambda_{V_0}^{-1} )_{ \ls[22] }
+ 2 {  \sqrt { D-1}    \over D }  ( \Lambda_{V_0}^{ -1})_{ \ls[12]} \cr 
 && + { D-1 \over D } (  \Lambda_{ H}^{ -1} )_{ \ls[11]}  
+ ( \Lambda_{ H}^{ -1} )_{ \ls[22]} { ( D-1) \over D }  + D^{-1} ( D-1) ( D-2)  
( \Lambda_{ H}^{ -1} )_{ \ls[33]} \cr 
&& + 2 { ( D-1) \over D } ( \Lambda_{ H}^{ -1} )_{ \ls[12]}  
 + 2 { ( \Lambda_H^{ -1} )_{ \ls[13] }  \over  D  } ( D -1  ) \sqrt { ( D -2 )}  
  + 2  { ( \Lambda_H^{ -1} )_{ \ls[23] }  \over  D   } ( D -1  ) \sqrt { ( D -2 )}  \biggr ) \cr 
&& + { 1 \over D^2} \left (  \widetilde \mu_{ 1}   
+ { \sqrt { ( D-1) }   }  ~  \widetilde \mu_{2}  \right )^3  \, . 
\cr  
&&
\eea

\subsection{ A 2-node case $ \sum_{ i , j  } \langle  M_{ij}^3 \rangle  $ } 

Using \ref{cubicWick} we have 
\bea\label{Mijcubed}  
\langle M_{ ij}^3 \rangle 
= \sum_{ i , j } 3 \langle M_{ ij }^2 \rangle_{\con} \langle M_{ ij} \rangle + 
\sum_{ i , j } \langle M_{ ij}\rangle^{ 3 }  \, . 
\eea
Calculating this requires doing a few sums, which can be done by hand or with Mathematica (the function 
KoneckerDelta is handy). 
\bea\label{3sums}  
&& \sum_{ i , j }  F ( i , j )  = 0 \cr 
&& \sum_{ i , j } ( F ( i , j ) )^2  = ( D-1) \cr 
&& \sum_{ i , j }  ( F ( i , j ) )^3   = D^{-1} ( D -1) ( D -2) \, . 
\eea
Using (\ref{1ptanswerMij}),  we find for the second term in (\ref{Mijcubed})
\bea 
\sum_{ i , j } \langle M_{ ij} \rangle^3 = { \wmuo^3\over D } + { 3 \over D }  ~  \wmuo \wmut^2  
+ { ( D-2) \over D \sqrt { D-1} } \wmut^3 \, . 
\eea
For the first term on the RHS of (\ref{Mijcubed})
\bea\label{prodcontrib}  
\sum_{ i ,j } \langle M_{ ij} \rangle_{ \con} \langle M_{ ij} \rangle 
= { 3 \wmuo \over D } \sum_{ i , j } \langle M_{ ij}^2  \rangle_{ \con }
+ { 3 \wmut \over \sqrt{ D-1} } \sum_{ i , j } \langle M_{ ij}^2 \rangle_{ \con } ~ F ( i , j ) \, . 
\eea
The first term in (\ref{prodcontrib}) can be expressed as a function of the parameters of the Gaussian model using 
(\ref{resultMijMijcon}). The second term is calculated by specialising the fundamental quadratic 
moments (\ref{Master1pt1}) and doing the resulting sums over the $F$-factors. Consider  the $V_0$ contributions to the second term above. The term proportional to $  ( \Lambda_{ V_0}^{-1} )_{ \ls[11] } $ 
vanishes due to the first of  (\ref{3sums}). The $(22) $ contribution, using the third of (\ref{3sums})
 is
 \bea 
 { 3 \wmut \over \sqrt{ D -1 } } \times 
  { D -2 \over D } ( \Lambda_{V_0}^{ -1}  )_{ \ls[22] }  = 3 \wmut ( \Lambda_{V_0}^{ -1}  )_{ \ls[22] }
  { ( D -2 )  \over D  \sqrt { D -1 } } \, . 
 \eea 
 The $(12) $ contributions, using the second of (\ref{3sums}) is 
 \bea
  { 3 \wmut \over \sqrt{ D -1 } } \times   { 2 \sqrt { D -1} \over D }   ( \Lambda_{V_0}^{ -1})_{ \ls[12] }
  =  {   6 \wmut ( \Lambda_{V_0}^{ -1})_{ \ls[12] }  \over D } \, . 
 \eea 
Now consider the $V_H$ contribution to the second term in (\ref{prodcontrib}). 
The $ ( \Lambda_{ H}^{ -1 })_{ \ls[11] } $ term is
\bea 
{ 1 \over D }  ( \Lambda_{ H}^{ -1 })_{ \ls[11] } \sum_{ i , j } F ( i , j ) F ( j , j ) 
 = 0 \, . 
\eea
The $  ( \Lambda_{ H}^{ -1 })_{ \ls[22] } $ contribution is similarly zero. 
The $ (\Lambda_H^{ -1} )_{ \ls[33] } $ contribution  is 
\bea 
&& {3 \wmut   \over  \sqrt { D -1 } }
 \times { D (\Lambda_H^{ -1} )_{ \ls[33] } \over ( D-2) }  \times \sum_{ i , j , p , q } 
 F ( i, p ) F ( i , q ) F ( j , p ) F ( j , q ) F ( i , j ) F ( p, q ) \cr 
 && =  3 \wmut  (\Lambda_H^{ -1} )_{ \ls[33] } \sqrt { D -1 }  { ( D -3 ) \over D } \, . 
\eea
The sum of products of six $F$'s  
is readily done with  Mathematica  to give $ D^{ -2} ( D -1) ( D -2 ) (D-3)  $. 

Contributions from  the $ ( 1,2)  $ matrix element of symmetric matrix $ \Lambda_{ H}$ give 
\bea 
 { 6 \wmut \sqrt { ( D -1 ) }  \over D } ( \Lambda_H^{ -1 } )_{ \ls[12] } \, . 
\eea
This uses the second of (\ref{3sums}).
From $ ( 1,3 ) $ and $ ( 3,1) $ we have 
\bea 
&& {3  \wmut \over  \sqrt{ D -1 } }  \times  { \Lambda_H^{-1} \over \sqrt{ D -2 } } 
\times 2 \sum_{ i , j , p } F ( i , p ) (F ( j,p ))^2 F ( i , j ) \cr 
&& = {3  \wmut \over  \sqrt{ D -1 } }  \times  { \Lambda_H^{-1} \over \sqrt{ D -2 } } 
\times 2 D^{ -1}  ( D -1 ) ( D -2 ) \cr 
&& = { 6 \wmut \over D }  ( \Lambda_{ H }^{-1} )_{ \ls [13 ] } \sqrt{ ( D -1 ) ( D -2 )  } \, . 
\eea
From $ ( 2,3 ) $ and $ ( 3,2 )$, we have 
\bea 
{ 6 \wmut  \over D }  \sqrt{ ( D -1 ) ( D -2 ) }   (  \Lambda_H^{-1} )_{ \ls[23] }  \, . 
\eea

Now consider the contribution from $V_2$. It is convenient to use (\ref{Master1})
\bea\label{V2cont}  
&& ( \Lambda_{ V_2}^{-1} ) \sum_{ i , j } \sum_{ a_1 , b_1 , c , a_2 , b_2 , d } C_{ a_1 i } C_{ b_1 j } C_{ a_1 b_1 c  }^{ H , H \rightarrow V_2 } 
C_{ a_2 i } C_{ b_2 j } C_{ a_2 , b_2 , c  }^{ H , H \rightarrow V_2 } C_{  d i } C_{ d  j } \cr 
&& = ( \Lambda_{ V_2}^{-1} ) 
\sum_{ i , j } \sum_{ a_1 , b_1 , c , a_2 , b_2 , d } 
C_{ a_1 i } C_{ a_2 i } C_{ d i } C_{ b_1 j } C_{ b_2 j } C_{ d j } 
( P^{ H , H \rightarrow V_2 } )_{ a_1 b_1 ; a_2 b_2 }  \cr 
&& = ( \Lambda_{ V_2}^{-1} )  \sum_{ a_1 , b_1 , c , a_2 , b_2 , d }  
C_{ a_1 a_2 d } C_{ b_1 b_2 d } ( P^{ H , H \rightarrow V_2 } )_{ a_1 b_1 ; a_2 b_2 }  \cr 
&& = ( \Lambda_{ V_2}^{-1} )   { D -2 \over D } 
\sum_{ a_1 , a_2 , b_1 , b_2 } ( P^{ H , H \rightarrow H } )_{ a_1 , a_2  ; b_1  , b_2 }
  ( P^{ H , H \rightarrow V_2 } )_{ a_1 b_1 ; a_2 b_2 } \, . 
\eea
In the last line, we have used the second equation in   \ref{normCs} which gives the relation between the invariant $ C_{ abc} $  in $V_{ \nat}^{ \otimes 3 } $ and the normalized 
Clebsch-Gordan  Gordan coefficients $ C_{ abc }^{ H H \rightarrow H}$, and the formula for the projector in terms of the Clebsch-Gordan coefficients.

The symmetric part of $ V_{  H } \otimes V_{ H } $, i.e the subspace invariant under the swop of the two factors, decomposes into irreducible representations of the diagonal $ S_D$ action as 
$ V_0 \oplus V_{ H } \oplus V_2$. This means that 
\bea
P^{ H , H \rightarrow V_2 } & = &    ( 1 - P^{ H , H \rightarrow H } - P^{ H , H \rightarrow V_0 } )    {  ( 1 + \tau ) \over 2 }  \cr 
 P^{ H , H \rightarrow H } {  ( 1 + \tau ) \over 2 }  & = &   P^{ H , H \rightarrow H }  \, . 
\eea
This means that 
\bea\label{PHHV2}  
( P^{ H , H \rightarrow V_2 } )_{ a_1 b_1 ; a_2 b_2 }  
= { 1 \over 2 } 
( \delta_{ a_1 a_2 } \delta_{ b_1 b_2 } +  \delta_{ a_1 b_2 } \delta_{  a_2 b_1 } ) 
- P^{ H , H \rightarrow H }_{ a_1 , b_1 ; a_2 , b_2 }  - P^{ H , H \rightarrow V_0 }_{ a_1 , b_1 ; a_2 , b_2 } \, . 
\eea
A useful fact following from (\ref{asumCC}) and (\ref{isumC}) is 
\bea\label{usfct}  
\sum_{ a } C_{ a a b } = 0   \, . 
\eea
When the expression (\ref{PHHV2})  is substituted in (\ref{V2cont})
the first term on the RHS of (\ref{PHHV2})  does not contribute because of 
(\ref{usfct}). The second term gives 
\bea\label{V2seciijsq} 
( \Lambda_{ V_2}^{-1} )   { D -2 \over  2 D }
\sum_{ a_1 , a_2 , b_1 , b_2 } ( P^{ H , H \rightarrow H } )_{ a_1 , a_2  ; a_2 , a_1  }    
=  ( \Lambda_{ V_2}^{-1} )   { ( D -1 ) ( D -2 ) \over 2 D }  \, . 
\eea 
The third term gives 
\bea\label{V2lastiijsq} 
&& -  ( \Lambda_{ V_2}^{-1} )  { D -2 \over D }   \sum_{ a_1 , a_2 , b_1 , b_2 } ( P^{ H , H \rightarrow H } )_{ a_1 , a_2   ; b_1   , b_2 }
  ( P^{ H , H \rightarrow H  } )_{ a_1 b_1 ; a_2 b_2 } \cr 
  && =  - ( \Lambda_{ V_2}^{-1} )   { D \over ( D -2 } 
  \sum_{ a_1 , a_2 , b_1 , b_2 , c_1 , c_2  } 
  C_{ a_1 , a_2  , c_1 } C_{ b_1 , b_2 , c_1 } C_{ a_1 , b_1 , c_2  } C_{ a_2 , b_2 , c_2 }  \cr 
  && =  - ( \Lambda_{ V_2}^{-1} )   { D \over ( D -2)  }  
  \sum_{ a_1 , a_2 , b_1 , b_2 , c_1 , c_2 } \sum_{ i , j , p , q } 
  F ( i , p ) F ( i , q ) F ( i , j ) F ( j , p ) F ( j , q ) F ( p , q ) \cr 
  && = - ( \Lambda_{ V_2}^{-1} )   { D \over ( D -2 ) }   { ( D -1 ) ( D -2 )( D -3 ) \over D^2 } \cr 
  && = - ( \Lambda_{ V_2}^{-1} )   { ( D -1 ) ( D -3 ) \over D }  \, . 
\eea
The fourth term gives 
\bea\label{fourthterm} 
&& -  ( \Lambda_{ V_2}^{-1} )  { D -2 \over D } \sum_{ a_1 , a_2 , b_1 , b_2 } ( P^{ H , H \rightarrow H } )_{ a_1 , a_2   ; b_1   , b_2 } ( P^{ H , H \rightarrow V_0 } )_{ a_1 , b_1   ; a_2   , b_2 }  \cr 
&&  -  ( \Lambda_{ V_2}^{-1} )  { D -2 \over D }  { 1 \over  ( D -1 ) } 
 \sum_{ a_1 , a_2 , b_1 , b_2 } ( P^{ H , H \rightarrow H } )_{ a_1 , a_2   ; b_1   , b_2 } \delta_{ a_1 b_1 } \delta_{ a_2 b_2 } \cr 
 && =  -  ( \Lambda_{ V_2}^{-1} )  { D -2 \over D }  \, . 
\eea
Collecting terms from (\ref{V2seciijsq}), (\ref{V2lastiijsq}) and (\ref{fourthterm}) we get 
\bea 
 ( \Lambda_{ V_2}^{ -1} ) { ( 3 - D  )   \over 2 }  \, . 
\eea
Multiplying the factor ${ 3 \wmut  \over \sqrt{ D -1 } }$ from (\ref{prodcontrib})
 to get a contribution to $ \sum_{ i , j } \langle M_{ ij}^3 \rangle $, we get 
\bea 
  { 3 \wmut  \over \sqrt{ D -1 } } ~  ( \Lambda_{ V_2}^{ -1} ) { ( 3 - D  )   \over  2  }  
=  {  \wmut \Lambda_{ V_2}^{ -1} } {    3( 3 - D  )     \over 2   \sqrt { D -1 }  } \, . 
\eea

The contribution from $ V_3$  is 
\bea 
{ 3 \wmut  \over \sqrt{ D -1 } } 
( \Lambda_{ V_3 }^{-1} )   { D -2 \over D } 
\sum_{ a_1 , a_2 , b_1 , b_2 } ( P^{ H , H \rightarrow H } )_{ a_1 , a_2  ; b_1  , b_2 }
  ( P^{ H , H \rightarrow V_3 } )_{ a_1 b_1 ; a_2 b_2 } \, . 
\eea
Now use the fact that 
\bea 
( P^{ H , H \rightarrow V_3 } )_{ a_1 b_1 ; a_2 b_2 } 
= { 1 \over 2 } (  \delta_{ a_1 a_2 } \delta_{ b_1 b_2 }  - \delta_{ a_1 b_2 } \delta_{ a_2 b_1 } ) 
\eea
along with 
\bea 
&&  \sum_{ a , b } ( P^{ H , H \rightarrow H  } )_{ a a  ;  b b  } =   0 \cr 
&& \sum_{ a_1 , a_2 }  ( P^{ H , H \rightarrow H  } )_{ a_1 a_2  ;  a_2 a_1  }  = 
 \sum_{ a_1 , a_2 }  ( P^{ H , H \rightarrow H  } )_{ a_1 a_2  ;  a_1 a_2   } = ( D -1 ) 
\eea
to find 
\bea 
- ( \Lambda_{ V_3 }^{-1} )  { 3 \wmut  \over \sqrt{ D -1 } }   { ( D -2 ) \over D }  ( D -1 ) 
= - 3 ~ \wmut  ~ ( \Lambda_{ V_3 }^{-1} )  {  ( D -2 )  \sqrt { D -1 }  \over 2 D }  \, . 
\eea

Collecting all the contributions we have 
\bea 
&& \sum_{ i , j } \langle M_{ ij}^3 \rangle = { \wmuo^3\over D } + { 3 \over D }  ~  \wmuo \wmut^2  
+ { ( D-2) \over D \sqrt { D-1} } \wmut^3  \cr 
&& + { 3 \wmuo \over D } ~~ \biggl ( ( \Lambda_{V_0}^{-1})_{ \ls[11] }  +  ( \Lambda_{V_0}^{-1} )_{ \ls[22] }  
 + ( D -1 ) ( \Lambda_H^{-1} )_{ \ls[22]  } + ( D-1 )  ( \Lambda_{ H}^{ -1} )_{ \ls[33] } 
+  ( D -1 ) ( \Lambda_H^{-1} )_{ \ls[11]  }  \cr 
&& +   {  D  ( D - 3 ) \over 2 }  ( \Lambda_{V_2} )^{ -1} + { ( D-1)   ( D - 2  ) \over 2 } ( \Lambda_{V_3} )^{ -1}  \biggr )  + { 3 \wmut ( \Lambda_{ V_0}^{ -1}  )_{ \ls[22] } } { D -2 \over D \sqrt { D - 1 }  }  
 +  { 6 \wmut ( \Lambda_{V_0}^{ -1})_{ \ls[12] }   \over D }   \cr 
 &&   + 3 \wmut  (\Lambda_H^{ -1} )_{ \ls[33] } \sqrt { D -1 }   { ( D -3 ) \over D } 
+ 6 \wmut  ( \Lambda_{ H}^{-1} )_{ \ls[12 ] } { \sqrt { ( D -1 ) }  \over D } 
 + 6  \wmut  ( \Lambda_{ H}^{-1} )_{ \ls[13 ] } { \sqrt { ( D -1 )( D -2 )  }  \over D }   \cr 
&&  +  6  \wmut  ( \Lambda_{ H}^{-1} )_{ \ls[23 ] } { \sqrt { ( D -1 )( D -2 )  }  \over D }  
+   {  \wmut \Lambda_{ V_2}^{ -1} } {   3 ( 3 - D )      \over 2 \sqrt { D-1}     } 
  - 3 ~ \wmut  ~ ( \Lambda_{ V_3 }^{-1} )  {  ( D -2 )  \sqrt { D -1 }  \over 2 D }  \, . 
   \cr 
   && 
\eea

\subsection{ A 3-node case $ \sum_{ i, j , k } \langle  M_{ ij} M_{ jk } M_{ ki} \rangle  $ } 

Using Wick's theorem (\ref{cubicWick})
\bea 
&& \sum_{ i , j , k } \langle M_{ ij} M_{ jk} M_{ ki} \rangle 
= \sum_{ i , j , k  } \langle M_{ ij} \rangle \langle M_{ jk} \rangle \langle M_{ ki} \rangle 
+ \langle M_{ ij} M_{ jk} \rangle \langle M_{ ki} \rangle + \langle M_{ ij} M_{ ki} \rangle \langle M_{ jk} \rangle + \langle M_{ jk} M_{ ki} \rangle \langle M_{ ij} \rangle  \, . \cr 
&& 
\eea

The first term is 
\bea 
\sum_{ i , j ,k }
 \left (  {\wmuo \over D }  + { \wmut \over \sqrt { D -1 } } F ( i , j )    \right ) 
  \left (  {\wmuo \over D }  + { \wmut \over \sqrt { D -1 } } F ( j,  k )    \right )  
   \left (  {\wmuo \over D }  + { \wmut \over \sqrt { D -1 } } F (  k ,  i  )    \right )  \, . 
\eea
Using
\bea\label{FmatMult}  
 \sum_{ j } F ( i ,j ) F ( j , k ) = F ( i , k ) 
\eea
along with the first and second of (\ref{3sums}) 
we can  show that 
\bea 
&& \sum_{ i , j , k } \langle M_{ ij} \rangle \langle  M_{ jk} \rangle \langle  M_{ ki} \rangle  
= \wmuo^3 + { \wmut^3 \over \sqrt { D -1 } }  \, . 
\eea
Consider the remaining three terms. Focus on the first of these : 
\bea
 && \sum_{ i , j , k } \langle M_{ ij} M_{ jk} \rangle_{ \con} ~  \langle M_{ k i } \rangle  \cr 
 && = \sum_{ i , j , k } \langle M_{ ij} M_{ jk} \rangle_{ \con} \left (  { \wmuo \over D } + { \wmut \over \sqrt { D -1 } } F ( k , i ) \right ) \cr 
&& = { \wmuo \over D } \sum_{ i , j , k  } \langle M_{ ij} M_{ jk} \rangle_{ \con} 
+ { \wmut \over \sqrt { D -1 } } \sum_{ i , j , k } \langle M_{ ij} M_{ jk} \rangle F ( k , i )  \, . 
\eea
We already know the first term  from (\ref{resMijMjk}).
So let us consider the second. An easy calculation using (\ref{Master1}) (or  equivalently using (\ref{Master1pt1})) shows that the  contribution from the 
$V_0$ channel is 
\bea 
{ \wmut  \over \sqrt { D -1 } }   (\Lambda_{V_0}^{-1} )_{ \ls[22] }   \, . 
\eea
From the $ V_H $ channel, the contributions are 
\bea 
\wmut \sqrt { D -1 }    ( \Lambda_H^{-1} )_{ \ls[33] }  +  \wmut  \sqrt { D -1 } ( \Lambda^{-1}_{ H} )_{ \ls[12] }  \, . 
\eea 
From the $V_2$ channel, we get 
\bea 
\wmut \Lambda_{V_2}^{-1} { D ( D -3 ) \over 2 \sqrt { D -1 } } -  \wmut ( \Lambda_{V_3}^{-1} ) { D -2 \sqrt { D-1 } \over 2 }   \, . 
\eea

Collecting terms 
\bea 
&& \sum_{ i , j , k } \langle M_{ ij}  M_{ jk} \rangle_{ \con}  \langle  M_{ ki} \rangle  
= \wmuo  \left ( ( \Lambda_{V_0}^{ -1})_{ \ls[11]}   +  ( D-1)   ( \Lambda_{ H}^{-1} )_{ \ls[12] } \right )  \cr 
&&  + { \wmut \sqrt { D -1 } }   (\Lambda_{V_0}^{-1} )_{ \ls[22] } + 
\wmut \sqrt { D -1 }    ( \Lambda_H^{-1} )_{ \ls[33] }  +  \wmut \sqrt { D -1 } ( \Lambda^{-1}_{ H} )_{ \ls[12] }  \cr 
 && + \wmut \Lambda_{V_2}^{-1} { D ( D -3 ) \over 2 \sqrt { D -1 } } - \wmut ( \Lambda_{V_3}^{-1} ) { ( D -2 )  \sqrt { D-1 } \over 2 }  \, .
\eea

By relabelling indices, it is easy to see that 
\bea 
\sum_{ i , j , k } \langle M_{ ij} M_{ jk} \rangle_{ \con}  \langle M_{ ki} \rangle  = 
\sum_{ i , j , k } \langle M_{ ij} M_{ ki} \rangle_{ \con}  \langle M_{ jk} \rangle  = 
\sum_{ i , j , k }  \langle M_{ jk} M_{ ki} \rangle_{ \con}  \langle M_{ ij} \rangle
\eea 

Hence, we have 
\bea 
&& \sum_{ i , j , k } \langle M_{ ij} M_{ jk} M_{ ki} \rangle  
= \wmuo^3 + { \wmut^3 \over \sqrt { D -1 } }    + 
3  \wmuo  \left ( ( \Lambda_{V_0}^{ -1})_{ \ls[11]}   +  ( D-1)   ( \Lambda_{ H}^{-1} )_{ \ls[12] } \right )  \cr 
&&  + 3  { \wmut \over  \sqrt { D -1 } }   (\Lambda_{V_0}^{-1} )_{ \ls[22] } + 
3 \wmut \sqrt { D -1 }    ( \Lambda_H^{-1} )_{ \ls[33] }  + 3  \wmut \sqrt { D -1 } ( \Lambda^{-1}_{ H} )_{ \ls[12] }  \cr 
 && + 3 \wmut \Lambda_{V_2}^{-1} { D ( D -3 ) \over 2 \sqrt { D -1 } } -  3 \wmut ( \Lambda_{V_3}^{-1} ) { ( D -2 )  \sqrt { D-1 } \over 2 } \, . 
\eea

\subsection{ A 6-node case $ \sum_{ i_1 , \cdots , i_6 } \langle M_{ i_1 i_2} M_{ i_3 i_4} M_{ i_5 i_6} \rangle $ } 

The sums over $i_1 , \cdots , i_6$ project to the $V_0$ representations. As a result, using 
(\ref{cubicWick}), along with (\ref{expMS}),   we have 
\bea 
&& \sum_{ i_1 , \cdots , i_6 }  \langle M_{ i_1 i_2} M_{ i_3 i_4} M_{ i_5 i_6} \rangle 
= { 3\over D^3} \sum_{ i_1 , \cdots , i_6 } 
 \langle S^{ V_0 ; 1 } S^{ V_0 ; 1 } \rangle  ~ \langle S^{ V_0 ; 1 } \rangle 
  + { 1 \over D^3 }  \sum_{ i_1 , \cdots , i_6 } ( \langle S^{ V_0 ; 1} \rangle )^3  \cr 
  &&  = { 3 \wmuo  D^3 }  ( \Lambda_{ V_0}^{ -1} )_{ \ls[11]}  + \wmuo^3 D^3  \, . 
\eea

\section{ A selection of quartic expectation values }\label{sec:quartic}

The methods we have used to calculate the cubic expectation values, which were explained 
in detail above, extend straightforwardly to quartic expectation values. 
The first step is to use Wick's theorem \ref{quarticWick}. Then we use 
the formulae for quadratic and linear expectation values from  Sections \ref{sec:IntroPIMT}
and \ref{sec:reptograph}. In order to arrive at the final result 
as a function of $ D , \wmuo, \wmut , \Lambda_{ V_0 } ,\Lambda_{ H } , \Lambda_{ V_2} , \Lambda_{ V_3 } $
we have to do certain sums over products of the natural-to-Hook projector $F ( i , j )$. 
We will give some formulae below to illustrate these steps for the quartic case, without 
producing detailed formulae as in previous sections.

\subsection{ A 2-node quartic expectation value  $ \sum_{ i ,j } \langle M_{ ij}^4 \rangle $  }

\bea 
\sum_{ i , j } \langle M_{ ij}^4 \rangle 
= \sum_{ i , j } \langle M_{ij} \rangle^4 + 6 \sum_{ i , j } \langle M_{ ij}^2 \rangle_{ \con } \langle M_{ij} \rangle^2  + 3 \sum_{ i , j  } \langle M_{ ij}^2 \rangle_{ \con} \langle M_{ ij}^2 \rangle_{ \con} 
\, .  
\eea

The quadratic average is 
\bea 
&& \langle M_{ i,j}^2 \rangle_{\con} 
= { ( \Lambda_{V_0}^{-1} )_{\ls[11]}  \over D^2} + { ( \Lambda_{V_0}^{-1} )_{\ls[22]}  \over ( D-1)^2 } F (i,j)^2 
+ { 2 ( \Lambda_{V_0}^{-1})_{ \ls[12]}  \over D \sqrt { D-1} }  F ( i , j ) \cr 
&& + { ( \Lambda_{ H}^{-1} )_{ \ls[11]}  \over D } F ( j , j ) + { ( \Lambda_{H}^{-1} )_{ \ls[22]}  \over D } F ( i , i ) 
 + {  ( \Lambda_{ H}^{-1} )_{ \ls[33]}  \over D -2  }
\sum_{ p , q } F ( i , p ) F ( j , q )) F ( i , q ) F ( j , q ) F ( p , q )   \cr 
&& + { 2 ( \Lambda_{ H}^{-1} )_{ \ls[12]}  \over D } F ( i , j ) 
+  { 2 ( \Lambda_{ H}^{-1} )_{ \ls[13]}  \over D -2  } \sum_{ p } F ( i , p ) F ( j , p ) F ( j , p ) \cr 
&& + { 2 ( \Lambda_{ H}^{-1} )_{ \ls[23]}  \over \sqrt { D -2}   }
 \sum_{ p } F ( i , p ) F ( j , p ) F ( j , p )  \cr 
 && + (\Lambda_{ V_2})^{-1} \left ( { 1 \over 2 } F ( i , i ) F ( j , j ) 
  + { 1 \over 2 } F ( i , j ) F ( i , j ) 
   - { D \over D-2} \sum_{ p , q } F ( i , p ) F ( i , q ) F ( j , p ) F ( j , q ) F ( p  , q )   \right ) 
   \cr 
   && + { ( \Lambda_{ V_3}^{-1} ) \over 2 }   \left ( { 1 \over 2 } F ( i , i ) F ( j , j ) - F( i , j ) F ( i, j ) \right )  \, . 
\eea

Using this and 
\bea 
 \langle M_{ ij} \rangle  = 
 \left (  {\wmuo \over D }  + { \wmut \over \sqrt { D -1 } } F ( i , j )    \right )  \, . 
\eea
we can work out the formula for $ \sum_{ i , j } \langle M_{ ij}^4 \rangle $ as a function 
of the $13$ Gaussian model parameters. Mathematica would be handy in doing the sums over products 
 of $F ( i, j)$ which arise.

\subsection{ A $5$-node quartic expectation value   $ \sum_{ i , j , k , p , q } \langle M_{ ij} M_{ jk} M_{ kp } M_{ pq } \rangle $  } 

From \ref{quarticWick}, we have 
\bea 
&& \sum_{ i , j , k , p , q } \langle M_{ ij} M_{ jk} M_{ kp } M_{ pq } \rangle  
 = \sum_{ i , j , k , p , q }  \langle M_{ ij} \rangle \langle M_{ jk} \rangle \langle M_{ kp} \rangle 
 \langle M_{ pq} \rangle \cr 
 && + \sum  \langle M_{ ij} M_{ jk} \rangle_{ \con}  \langle M_{ kp} \rangle \langle M_{ pq} \rangle 
+ \sum \langle M_{ ij} M_{ kp } \rangle_{ \con}   \langle M_{ jk } \rangle \langle M_{ pq} \rangle 
 + \sum \langle M_{ ij} M_{ pq } \rangle_{ \con}   \langle M_{ jk } \rangle \langle M_{ kp} \rangle  \cr 
&& + \sum \langle M_{ jk}  M_{ kp } \rangle_{ \con}   \langle M_{ ij  } \rangle \langle M_{ pq} \rangle   
 + \sum \langle M_{ jk } M_{ pq } \rangle_{ \con}   \langle M_{ ij  } \rangle \langle M_{ kp } \rangle  
+ \sum \langle M_{ kp } M_{ pq  } \rangle_{ \con}   \langle M_{ ij  } \rangle \langle M_{ jk } \rangle  \cr 
&& + \sum  \langle M_{ ij} M_{ jk} \rangle_{ \con}  \langle M_{ kp} M_{ pq} \rangle_{ \con}  
+ \sum \langle M_{ ij} M_{ kp } \rangle_{ \con}   \langle M_{ jk } M_{ pq} \rangle_{ \con}  
 + \sum \langle M_{ ij} M_{ pq } \rangle_{ \con}   \langle M_{ jk }  M_{ kp} \rangle_{ \con} \, .  \cr 
 && 
\eea
All the summands on the RHS can be evaluated using \ref{Master1} or \ref{Master1pt1} 
in terms of $F ( i , j )$. The sums can be done with the help of Mathematica
to  obtain expressions in terms of $ D , \wmuo , \wmut, \Lambda_{V}$.

\section{ Summary and  Outlook }

We have used the representation theory of symmetric groups  $S_D$ in order to define a 13-parameter permutation invariant 
Gaussian matrix  model, to compute the expectation values of all the graph-basis  permutation invariant quadratic functions of the random matrix, and a selection of cubic and quartic invariants. 
In \cite{LMT} analogous computations with a 5-parameter model were compared with matrix data constructed from a corpus of the English language. A natural direction is to extend that discussion of the English language, or indeed other languages,  to the present 
13-parameter model. Combining the experimental methods employed in \cite{LMT} with machine learning methods such those used in \cite{HJN1807}, in  the investigation of the 13-parameter model,  would also be interesting to explore. 

As a theoretical extension of the present work, it will be useful to generalise 
the representation theoretic parametrisation of the Gaussian models to 
perturbations of the Gaussian model, where we add cubic and quartic terms to 
the Gaussian action. Identifying  parameter spaces of these deformations  which allow well-defined 
convergent partition functions and expectation values will be useful for eventual comparison to data.
If we ignore the convergence constraints, the general  perturbed model at cubic and quartic order has 
 $348$ parameters, since there are 52 cubic invariants and 296 quartic invariants (Appendix A of \cite{LMT}). As in the Gaussian case, we can expect that representation theory methods will be useful   
 in handling this more general problem.  Further techniques involving partition algebras underlying 
 the representation theory of tensor products of the natural representation will likely play a role 
 (see e.g. \cite{BenHal} for recent work in these directions).

It is worth noting that permutation invariant random matrix distributions have been approached from a different perspective, based on non-commutative probability theory \cite{FG1503,FG1507,ACDGM1805}. 
The approach of the present paper and \cite{LMT} is based on the connection between 
statistical physics and zero dimensional quantum field theory (QFT). It would seem that the approach of the present paper can complement the theory developed 
in these papers \cite{FG1503,FG1507,ACDGM1805} by producing integral representations (Gaussians or perturbed Gaussians) of random matrix distributions having finite expectation values for permutation 
invariant polynomial functions of matrices. The results on   the central limit theorem from the above references would be very interesting to interpret from  the  present QFT perspective. 

The computation of expectation values in Gaussian matrix models admits 
generalization to higher tensors. Indeed the motivating framework in computational linguistics 
discussed in \cite{LMT} involves matrices as well as higher tensors in a natural way. 
Generalizations of the present work on representation theoretic parametrisation of 
Gaussian models and computation of graph-basis observables to the tensor case is an interesting 
avenue for future research. 

In this paper, we have focused on the explicit computation of permutation invariant correlators
for general $D$. Some simplifications at large $D$ were discussed in section \ref{sec:largeD}. 
For traditional matrix models having $U(D)$ ( or $SO(D)/ Sp(D)$ symmetries), there is a rich geometry of two dimensional surfaces and maps in the large $D$ expansions which allows these expansions of matrix quantum field theories to have deep connections to string theory \cite{tHooft,Maldacena}. 
It will be interesting to explore  the possibility of  two dimensional geometrical interpretations of the large $D$ expansion in permutation invariant matrix models.

\vskip.5cm

 \begin{center} 
 { \bf Acknowledgements} 
 \end{center} 
This research is supported by the STFC consolidated grant ST/L000415/1 ``String Theory, Gauge Theory \& Duality'' and  a Visiting Professorship at the University of the Witwatersrand, funded by a Simons Foundation grant to the Mandelstam Institute for Theoretical Physics.  I  thank the Galileo Galilei Institute for Theoretical Physics for  hospitality and the INFN for partial support during the completion of this work. I also thank the organizers of the  workshop on Matrix Models for non-commutative geometry and string theory in Vienna, the KEK theory group, Tsukuba, Japan and the Yukawa Instititute for Theoretical Physics (YITP), Kyoto, Japan,  for hospitality  during the completion of this project.
The revised version on the arXiv has benefitted from helpful comments on the presentation and typos spotted by the Nuclear Physics B anonymous referee, as well as a few other typos spotted during  the preparation of 
\cite{GTMDS}.
I am grateful for conversations on the subject of this paper 
to David Arrowsmith, Masayuki Asahara, Robert de Mello Koch, Yang Hui He, Christopher  Hull, Satoshi Iso,  Vishnu Jejjala, Dimitrios Kartsaklis, Mehrnoosh Sadrzadeh, Shotaro Shiba, Lewis Sword.

\begin{appendix} 

\section{Multi-dimensional Gaussian Integrals  and Wick's theorem  } \label{app:Wick}  

Consider the multi-variable integral with a Gaussian integrand  
\bea\label{FundamentalQFT} 
\cZ = \int dx \exp \left (  - {  1\over 2 } \sum_{ i , j=1 }^N  x_i A_{ij}  x_j   + \sum_{  i } s_i x_i   \right )
= \sqrt  {   ( 2 \pi)^N \over \det A  }  \exp \left ( { 1 \over 2 } s_{i}  ( A^{ -1 } )_{ ij}  s_j  \right ) \, . 
\eea
$x \in \mR^N $. $A \in \mC^{ N \times N } $ is a  real symmetric  positive definite 
matrix. $ s \in \mR^N $ is an arbitrary complex vector (see for example \cite{Gaussian-Zhang}, \cite{WikiCommonInteg}, Appendix A,  Equations (8) and (9) of  \cite{Zee}). One can also consider $A$ more generally  to be complex with positive definite real part, but 
to keep a probabilistic interpretation we keep $A$ real symmetric. 
Expectation values of functions $f(x)$  are defined by 
\bea 
\langle f(x) \rangle = { 1 \over \cZ } 
 \int dx ~ f (  x ) ~ \exp \left (  - {  1\over 2 } \sum_{ i , j=1 }^N  x_i A_{ij}  x_j   + \sum_{  i } s_i x_i   \right )   \, . 
\eea
These expectation values can be calculated by taking derivatives with respect to $s_i$ on both sides of 
(\ref{FundamentalQFT}). For the $x$ variables
\bea 
\langle x_i \rangle =  \sum_{ j } ( A^{ -1})_{ij} ~  s_j = \sum_{ j } s_j ~ ( A^{ -1})_{ji }   \, . 
\eea
Application of this equation, along with the formula for $dM$ in terms of the 
representation theoretic $S$-variables (\ref{dSmeas1}) leads to (\ref{V0tad}),(\ref{H23tad}). 
For expectation values of quadratic monomials we have 
\bea\label{quadWick1}  
\langle x_i x_j \rangle = (A^{ -1})_{ij} + \langle x_i \rangle \langle x_j \rangle  \, . 
\eea
We define the connected part as 
\bea\label{quadWick2}  
\langle x_i x_j \rangle_{ \ls[\con] } \equiv 
  \langle x_i x_j \rangle - \langle x_i \rangle \langle x_j \rangle  = (A^{ -1})_{ij} \, . 
\eea
The expressions (\ref{QDExp1}) and (\ref{QDExp2}) follow from these. 

For cubic expressions 
\bea\label{cubicWick}  
\langle x_i x_j x_k \rangle  = \langle x_i x_j \rangle_{ \con}  \langle x_k \rangle + 
\langle x_i x_k \rangle_{ \con}  \langle x_j \rangle + \langle  x_j x_k  \rangle_{ \con}  \langle x_i  \rangle + \langle x_i \rangle \langle x_j  \rangle \langle x_k  \rangle \, . 
\eea
For quartic expressions 
\bea\label{quarticWick}  
\langle x_i x_j x_k x_l \rangle && = \langle x_i x_j  \rangle_{ \con }  \langle x_k x_l \rangle_{ \con } 
+   \langle x_i x_k  \rangle_{ \con }  \langle x_j  x_l \rangle_{ \con }  + 
 \langle x_i x_l  \rangle_{ \con }  \langle x_j  x_k  \rangle_{ \con }  \cr 
 && +  \langle x_i x_j  \rangle_{ \con } \langle x_k  \rangle \langle x_l \rangle 
      +  \langle x_i x_k  \rangle_{ \con }  \langle x_j  \rangle \langle  x_l   \rangle 
       +   \langle x_i x_l  \rangle_{ \con }  \langle x_j \rangle \langle  x_k  \rangle \cr 
&&        +  \langle x_j x_k  \rangle_{ \con } \langle x_i  \rangle \langle x_l \rangle  
      +  \langle x_j x_l  \rangle_{ \con }  \langle x_i  \rangle \langle  x_k  \rangle 
       +   \langle x_k x_l  \rangle_{ \con }  \langle x_i \rangle \langle  x_j  \rangle \cr 
       && 
       + \langle x_i \rangle   \langle x_j \rangle \langle  x_k \rangle \langle x_l \rangle  
       \, . \cr 
       && 
\eea
These illustrate a general fact (known as Wick's theorem in the quantum field theory context 
and Isserlis' theorem in probablity theory \cite{Isserlis}) 
 about Gaussian expectation values. 
Higher order expectation values can be expressed in terms of linear and quadratic expectation values. 
When applied to permutation invariant matrix models, we still have non-trivial sums left to do, after 
Wick's theorem has been applied. This is illustrated in the calculations of section \ref{sec:cubic} and section \ref{sec:quartic}.  

\section{ Rep theory of $V_H$ and its tensor products  }\label{repVH} 

Some basics of rep theory of $V_H$ can be presented in a self-contained way, assuming only knowledge of linear algebra and index notation. 

Alternatively, we can observe that the matrices in $V_H$ are the same as 
Young's orthogonal basis. If we just follow the self-contained route, we define 
\bea\label{defDH} 
D^H_{ ab} ( \sigma )&&  = ( E_a  , \sigma E_b ) \cr 
&& = \sum_{ i } \sum_{j } C_{ a , i } C_{ b , j} ( e_i , e_j ) \cr 
&& = \sum_{ i , j } C_{ a , i} C_{ b , j} ( e_i , e_{ \sigma^{-1} ( j) } ) \cr 
&& = \sum_{ i , j } C_{ a , i } C_{ b , j} \delta_{ i , \sigma^{-1} ( j ) } \cr 
&& = \sum_{i } C_{ a, i } C_{ b , \sigma (i) }  \, . 
\eea 

We have  the orthogonality property : 
\bea 
D^{ H}_{ ab} ( \sigma^{-1} )&&  = \sum_{ i } C_{ a , i } C_{ b  , \sigma^{-1} ( i ) } \cr 
&& = \sum_{ i } C_{ a , \sigma (i) } C_{ b , i } \cr 
&& = D^{ H}_{ ba} ( \sigma )  \, . 
\eea

The homomorphism property  
\bea 
D^{ H}_{ab} ( \sigma ) D^H_{ bc } ( \tau ) 
&&  = \sum_{ i } C_{ a, i  } C_{ b , \sigma(i) } \sum_{ j } C_{ b ,j } C_{ c , \tau(j ) } \cr 
 && = \sum_{ i , j } C_{ a, i} C_{ b , j } C_{ b , \sigma ( i ) } C_{ c , \tau (j) } \cr 
 && = \sum_{ i , j } C_{ a , i } C_{ b , j } C_{ b , \sigma (i) } C_{ c , \tau (j) } \cr 
 && = \sum_{ i , j } C_{ a , i } \left ( \delta_{ j , \sigma(i) } - { 1 \over D } \right ) C_{ c , \tau (j) } \cr 
 && = \sum_{ i , j } C_{ a , \sigma^{-1} (j) } C_{ c , \tau (j) } - { 1 \over D } \sum_{ i , j } C_{ a, i } C_{ c , \tau (j) } \cr 
 && = \sum_{ j } C_{ a , j } C_{ c, \tau ( \sigma ( j ) ) } \cr 
 && = \sum_{ j } C_{ a , j } C_{ c,  \sigma \tau  ( j )  } \cr 
 && = D^H_{ ac} ( \sigma \tau )  \, . 
\eea

We used 
\bea 
\sum_{ a } C_{ a ,  i } C_{ a  , j } && = \left (  \delta_{ i j } - { 1 \over D }\right )  \cr 
\sum_{ a } C_{ a  , i } && = 0  \, . 
\eea

Using the definition  (\ref{defDH}), we prove 
\bea
\sum_{ b } D^{H }_{ b a  } ( \sigma ) C_{ b  , i } = C_{a , \sigma (i) }  \, . 
\eea
Indeed 
\bea 
&& \sum_{  b }  D^{H }_{ b a  } ( \sigma ) C_{ b ,  i } = \sum_{ j } C_{ b , j } C_{ a , \sigma (j) } C_{ b , i } \cr 
&& = \sum_{ j } C_{ a , \sigma ( j ) } ( \delta_{ i , j } - { 1 \over D } ) \cr 
&& = C_{ a , \sigma (i) }  \, . 
\eea
It is useful to define $ C_{ \sigma ( a ) , i } = \sum_{ b } D^{H }_{ b a  } ( \sigma ) C_{ b i } $ 
so the above can be expressed as an {\bf equivariance} property
\bea 
C_{ \sigma ( a ) , i } = C_{ a , \sigma (i) } \, . 
\eea
which is an equivariance condition for the map $ V_H \rightarrow V_{ nat } $ given by 
the  coefficients $ C_{ a , i } $. This map intertwines the $S_n$ action on the 
$V_H$ and $ V_{ nat } $. 
Now define $C_{ a ,  b ,  c } $ 
\bea 
C_{ a ,  b ,  c } = \sum_{ i } C_{ a  ,  i } C_{ b ,  i } C_{ c  ,  i }  \, . 
\eea  
We show that this is an {\bf invariant tensor}. 
\bea 
C_{ \sigma ( a ) , \sigma ( b ) , \sigma ( c ) }  = C_{ a  ,  b  ,   c }  \, . 
\eea
Indeed 
\bea 
&& C_{ \sigma ( a) , \sigma ( b)  , \sigma ( c ) } = \sum_{ i =1 }^n  C_{ \sigma ( a ) , i } C_{ \sigma (b)  , i } C_{ \sigma ( c ) , i } \cr 
&& = \sum_{ i } C_{ a , \sigma (i) } C_{ b , \sigma (i) } C_{ c , \sigma (i) } \cr 
&& = \sum_{ i } C_{ a , i } C_{ b , i } C_{ c , i } \cr 
&& = C_{ a , b , c }  \, . 
\eea
We used the equivariance of the $C$'s, then the relabelled the sum $ i \rightarrow \sigma (i) $. 

Using vectors $ \{ e_a \} $ spanning $ V_H$, we write a basis for  the tensor product 
$V_H \otimes V_H$: 
\bea 
  e_a \otimes e_b  \, . 
\eea
There is a subspace of $ V_H \otimes V_H$, which transforms as the irrep $V_H$. 
This is constructed using the invariant $3$-index tensor $C_{ a , b , c } $. 
The linear combinations 
\bea 
E_a = \sum_{ a , b, c }  C_{ a , b , c } e_b \otimes e_c 
\eea
span the subspace $V_H$ in the direct sum decomposition of $ V_H \otimes V_H$ (Equation (\ref{decompsq}))
under the diagonal action of $S_D$.  
To see this, we can write the diagonal action of $ \sigma \in S_D$
\bea 
&& \sigma E_a = \sum_{ a , b , c } C_{ a , b , c }  ( \sigma e_b)  \otimes ( \sigma e_c )  \cr 
&& = \sum_{ a , b , c } \sum_{b' , c'  } 
 C_{ a , b , c } D^{ H }_{ b' b} ( \sigma ) D^{ H}_{ c' c } ( \sigma ) 
( e_{ b'} \otimes e_{ c'} )  \cr 
&& = \sum_{ a , b , c } \sum_{ d , b' , c' } 
C_{ d  , b , c } D^{ H}_{   a d   } ( \sigma^{-1}  \sigma)   D^{ H }_{ b' b } ( \sigma ) D^H_{ c' c } ( \sigma )( e_{ b'} \otimes e_{ c'} )   \cr 
&& = \sum_{ a , b , c } \sum_{a' ,  d , b' , c' } 
C_{ d  , b , c } D^{ H}_{   aa' }( \sigma^{-1} )  D^H_{a' d   } (   \sigma)   D^{ H }_{ b' b } ( \sigma ) D^H_{ c' c } ( \sigma ) ( e_{ b'} \otimes e_{ c'} )   \cr 
&& = \sum_{ a , a' ,  b' , c' } C_{ \sigma^{-1} ( a' ) , \sigma^{-1} ( b') , \sigma^{-1} ( c' ) } 
D^{ H}_{ a' , a } ( \sigma ) ( e_{ b'} \otimes e_{ c'} ) \cr 
&& = \sum_{ a' } D^{ H}_{ a' , a } ( \sigma )\sum_{ a , b , c } C_{ a' , b' , c' } ( e_{ b'} \otimes e_{ c'} ) \cr 
&& = \sum_{ a' } D^H_{ a' a } ( \sigma ) E_{a'}  \, . 
\eea
showing that the transformation is indeed by the matrix $D^H$.

These vectors $E_a$ are orthogonal. It is useful to calculate the inner product
\bea\label{orthognormHHH} 
( E_{a_1 } , E_{ a_2}  ) &&  = \sum_{ b , c } C_{ a_1 , b , c } C_{ a_2 , b , c } \cr  
 && = \sum_{ i , j  } \sum_{  b , c } C_{ a_1 , i } C_{ b , i } C_{ c , i } 
 C_{ a_2 , j } C_{ b , j } C_{ c , j } \cr 
 && = \sum_{ i , j } C_{ a_1 , i } C_{ a_2 , j } 
 \left ( \delta_{ i, j } - { 1 \over D }   \right )  
 \left ( \delta_{ i, j } - { 1 \over D }   \right )  \cr 
 && = ( 1 - {2 \over D }  ) \sum_{ i } C_{ a_1 , i } C_{ a_2 , i } \cr 
 && =   { ( D -2 ) \over D } \delta_{ a_1 , a_2 }  \, . 
\eea
which will be useful in the next section. 

\section{ Clebsch-Gordan coefficients and normalizations }\label{normClebschs}  

The normalized Clebsch-Gordan coefficients for an orthonormal basis of a subspace 
of $ V_H \otimes V_H$  transforming as an irrep $V$  obey the condition 
\bea
\sum_{ a , b } C^{ H , H \rightarrow V}_{ a ,  b, c  }  C^{ H , H \rightarrow V}_{ a ,  b, c'  }
= \delta_{ c c'}  \, . 
\eea
This means that 
\bea 
\sum_{ a , b , c }  C^{ H , H \rightarrow  V}_{ a ,  b, c  }  C^{ H , H \rightarrow   V}_{ a ,  b, c  }
=  \Dim ~ V  \, . 
\eea

If instead we consider the invariant state in $ H \otimes H \otimes V$, normalized to one, then 
\bea 
\sum_{ a , b , c } \sum_{ a' , b' , c' } ( C^{ H, H , V }_{ a , b , c }  e_{ a  } \otimes e_b \otimes e_c ,
 C^{ H, H ,  V }_{ a' , b' , c' }  e_{ a'} \otimes e_{ b'} \otimes e_{ c'} ) 
= \sum_{ a , b , c }  ( C^{ H, H ,  V }_{ a , b , c } )^2  = 1 \, . 
\eea

The equivariance property of the map $ C^{ H , H \rightarrow V } $ is 
\bea 
 D^{ H \otimes H } ( \sigma \otimes \sigma ) \sum_{ a , b } C^{ H , H \rightarrow V}_{ a ,  b, c  } e_{ a } \otimes e_{ b} && = \sum_{ b' , c' } D^H_{ a' a} ( \sigma )  D^{ H}_{ b' b} ( \sigma ) C^{ H , H \rightarrow V}_{ a ,  b, c  } e_{ a' } \otimes e_{ b' }  \cr 
&& = \sum_{ c' } D^{ H }_{ c'  c } ( \sigma ) \sum_{ a' , b' } C^{ H , H \rightarrow V}_{ a' ,  b', c'  } e_{ a' } \otimes e_{ b' } \, . 
\eea
which means
\bea 
 \sum_{ b' , c' } D^H_{ a' a} ( \sigma )  D^{ H}_{ b' b} ( \sigma ) C^{ H , H \rightarrow V}_{ a ,  b, c  } = D^{ H }_{ c'  c } ( \sigma )  C^{ H , H \rightarrow V}_{ a ,  b, c'  } \, . 
\eea
Multiplying on both sides by $ D^H_{ c e  } ( \sigma^{-1}  ) $ and summing over $ c $, we have, after using $D^H_{ ab } ( \sigma^{ -1} ) = D^H_{ ba} ( \sigma )$  and relabelling indices
\bea 
\sum_{ b' , c' } D^H_{ a' a} ( \sigma )  D^{ H}_{ b' b} ( \sigma ) D^{ H}_{ c'c } ( \sigma )  C^{ H , H \rightarrow V}_{ a ,  b, c  } =  C^{ H , H \rightarrow V}_{ a ,  b, c  }  \, . 
\eea
This means that we can identify 
\bea 
C^{ H , H \rightarrow V}_{ a ,  b, c  } = \sqrt{ \Dim ~ V } ~   C^{ H , H , V }_{ a , b , c }  \, . 
\eea

We also know that 
\bea 
C_{ a , b ,  c} = \sum_{ i } C_{ a , i } C_{ b,  i } C_{ c ,  i }  \, . 
\eea
has the invariance property of $ C_{ a , b , c }^{ H, H , H }  $. Since there is a unique 
invariant state in $ V_H \otimes V_H \otimes V_H$, $C_{ abc} $ must be proportional to $  C_{ a , b , c }^{ H, H , H } $. We calculate 
\bea\label{normCabc}  
&& \sum_{ a , b , c } C_{ a , b , c} C_{ a , b  , c } = \sum_{  i , j } \sum_{ a , b , c } 
 C_{ a,   i } C_{ b  , i } C_{ c ,  i } C_{ a ,  j } C_{ b  , j  } C_{ c ,  j  }  \cr 
 && =  \sum_{ i , j } ( \delta_{ ij} - { 1 \over D } ) ( \delta_{ ij} - { 1 \over D } ) ( \delta_{ ij} - { 1 \over D } ) \cr 
 && = 1 \times D - { 3 \over D } \times D + { 3 \over D^2 } \times D - { 1 \over D^3 } \times D^2 \cr 
 && = D - 3 + 2 D^{-1} = D^{ -1} ( D -1 ) ( D - 2 )   \, .  
\eea
We can therefore identify 
\bea\label{normCs}  
C^{ H , H , H }_{ a , b , c } & &= \sqrt { D \over  ( D -1 ) ( D -2 ) } ~~ C_{ a , b ,  c } \cr 
C^{ H , H \rightarrow H }_{ a , b , c }  && =  \sqrt{ D \over  ( D -2 ) }  ~~ C_{ a , b ,c }  \, . 
\eea
The second equation is also consistent, as expected, with (\ref{orthognormHHH}). 
The projector for $V_H$ in $ V_H \otimes V_H$ is given in terms of the normalized Clebsch-Gordan coefficients $ C^{ H , H \rightarrow H }_{ a , b , c }$
as 
\bea 
P^{ H , H \rightarrow H}_{ ab ; cd } = \sum_{ d } 
C^{ H , H \rightarrow H }_{ a , b , e  }C^{ H , H \rightarrow H }_{ c,d , e  } \, . 
\eea

\end{appendix}


\begin{thebibliography}{} 

\bibitem{Harris} 
Z. Harris, ``Mathematical Structures of Language,''  Wiley, 1968

\bibitem{Firth} 
J.R.Firth, ``A Synopsis of Linguistic Theory 1930-1955,'' Studies in Linguistic Analysis, 1957. 

\bibitem{CSC2010} 
B. Coecke, M. Sadrzadeh, and S. Clark
``Mathematical Foundations for a Compositional Distributional Model of Meaning''. Lambek Festschrift. Linguistic. Analysis,36,345-384, 2010 


\bibitem{GS2015} E. Grefenstette and M. Sadrzadeh,  ``Concrete models and empirical evaluations for a categorical compositional distributional model of meaning.''
 Computational Linguistics, 41:71-118.


\bibitem{MCG2014} J. Maillard, S. Clark, and E. Grefenstette, 
``A type-driven tensor-based semantics for CCG,'' 
 In Proceedings of the Type Theory and Natural Language Semantics Workshop, EACL.
 
\bibitem{BBZ2014} M. Baroni, R. Bernardi, and R. Zamparelli, 
 ``Frege in space: A program of compositional distributional semantics. Linguistic Issues in Language Technology, 9.
 
 
\bibitem{KSP2012} D. Kartsaklis, M. Sadrzadeh, and S. Pulman. 
``A unified sentence space for categorical distributional-compositional semantics: Theory and experiments.''  In Proceedings of 24th International Conference on Computational Linguistics (COLING): Posters, pages 549-558, Mumbai, India.



\bibitem{LMT} 
  D.~Kartsaklis, S.~Ramgoolam and M.~Sadrzadeh,
  ``Linguistic Matrix Theory,''
  arXiv:1703.10252 [cs.CL], to appear in Annales de l'Institut Henri Poincare  D. 

\bibitem{LMTqpl} 
  D.~Kartsaklis, S.~Ramgoolam and M.~Sadrzadeh,
  ``Linguistic Matrix Theory,'' (short version) Quantum Physics and Logic Conference 2017,
 pdf available at  http://qpl.science.ru.nl/accepted.html; Video at https://www.youtube.com/watch?v=RAractFNESU

\bibitem{OEISmulti} 
 https://oeis.org/A052171
 
 \bibitem{Hamermesh} 
M. Hamermesh, ``Group theory and its application to physical problems,'' Dover 1962.
  

\bibitem{FulHar}
W. Fulton and J. Harris, ``Representation Theory: a first course'' Springer 2004.

\bibitem{ZeeGroup} 
A. Zee, ``Group Theory in a Nutshell for Physicists,'' Princeton University Press, 2016.  

\bibitem{NaimarkStern} 
M. A. Naimark, A. I. Stern, ``Theory of Group Representations,'' A series of Comprehensive Studies in Mathematics, 246. 

 \bibitem{FFPR}
  R.~de Mello Koch and S.~Ramgoolam,
  ``Free field primaries in general dimensions: Counting and construction with rings and modules,''
  JHEP {\bf 1808} (2018) 088
  doi:10.1007/JHEP08(2018)088
  [arXiv:1806.01085 [hep-th]].
  
 
 \bibitem{GTMDS} 
  S.~Ramgoolam, M.~Sadrzadeh and L.~Sword,
  ``Gaussianity and typicality in matrix distributional semantics,''
  arXiv:1912.10839 [hep-th].
  
\bibitem{Gaussian-Zhang}
Online notes by David Zhang:  
http://david-k-zhang.com/notes/gaussian-integrals.html

\bibitem{WikiCommonInteg} 
https://en.wikipedia.org/wiki/Common\_integrals\_in\_quantum\_field\_theory
  

\bibitem{HJN1807}
  Y.~H.~He, V.~Jejjala and B.~D.~Nelson,
  ``hep-th,''
  arXiv:1807.00735 [cs.CL].
  
\bibitem{BenHal}
  Georgia Benkart, Tom Halverson, ``Partition Algebras and the Invariant Theory of the Symmetric Group,'' 
 arXiv:1709.07751 [math.RT] 
 
\bibitem{FG1503} 
F. Gabriel, 
``Combinatorial theory of permutation-invariant random matrices I: partitions, geometry and renormalization,'' https://arxiv.org/abs/1503.02792  

\bibitem{FG1507} 
F. Gabriel, 
``Combinatorial theory of permutation-invariant random matrices II: cumulants, freeness and Levy processes,''
https://arxiv.org/abs/1507.02465
  
\bibitem{ACDGM1805} 
Benson Au, Guillaume Cébron, Antoine Dahlqvist, Franck Gabriel, Camille Male
 ``Large permutation invariant random matrices are asymptotically free over the diagonal,'' 
  https://arxiv.org/abs/1805.07045
  
\bibitem{traffic} Camille Male, ``Traffic distributions and independence: permutation invariant random matrices and the three notions of independence,''  https://arxiv.org/abs/1111.4662
 
\bibitem{Peskin} 
M. Peskin, D. V. Schroder, ``An introduction to quantum field theory,'' Taylor and Francis Group, 1995. 

\bibitem{Zee}
A. Zee, ``Quantum Field Theory in a nutshell,'' Princeton University Press, 2010.  

\bibitem{Isserlis} 
https://en.wikipedia.org/wiki/Isserlis\%27\_theorem

\bibitem{tHooft} 
  G.~'t Hooft,
  ``A Planar Diagram Theory for Strong Interactions,''
  Nucl.\ Phys.\ B {\bf 72} (1974) 461.
  doi:10.1016/0550-3213(74)90154-0

\bibitem{Maldacena} 
  J.~M.~Maldacena,
  ``The Large N limit of superconformal field theories and supergravity,''
  Int.\ J.\ Theor.\ Phys.\  {\bf 38} (1999) 1113
   [Adv.\ Theor.\ Math.\ Phys.\  {\bf 2} (1998) 231]
  doi:10.1023/A:1026654312961, 10.4310/ATMP.1998.v2.n2.a1
  [hep-th/9711200].


\end{thebibliography}
\end{document}